\documentclass[pra,superscriptaddress,twoside,showpacs]{revtex4}
\usepackage{bbm}
\usepackage{amsmath}
\usepackage{amssymb}
\usepackage{amsfonts}
\usepackage{amsthm}
\usepackage{dsfont}
\usepackage{bm,epic,eepic}
\usepackage{graphics}
\usepackage{epsfig}
\usepackage{pst-plot}
\usepackage{pstricks}
\usepackage{pst-node}
\usepackage{pst-coil}
\usepackage[mathscr]{euscript}

\newcommand{\half}{\frac{1}{2}}
\newcommand{\av}[1]{\langle #1 \rangle}
\newcommand{\BEQ}{\begin{eqnarray}}
\newcommand{\ENQ}{\end{eqnarray}}
\newcommand{\EEQ}{\end{eqnarray}}
\newcommand{\ket}[1]{ |  #1 \rangle}

\newcommand{\bra}[1]{ \langle  #1 |}
\renewcommand{\H}{\mathcal{H}}
\newcommand{\beq}{\begin{equation}}
\newcommand{\eeq}{\end{equation}}
\newcommand{\enq}{\end{equation}}
\newcommand{\1}{\mathds{1}}

\newcommand{\forget}[1]{}
\newcommand{\equivto}{\Longleftrightarrow}
\newcommand{\nn}{\nonumber}
\newcommand{\integer}[1]{\lceil #1 \rfloor}
\newcommand{\mbb}{\mathbb}

\newcommand{\C}{\ensuremath{\mbb{C}}}

\begin{document}
\title{Partial separability and entanglement criteria for multiqubit quantum states}
\date{\today}

\begin{abstract}

We explore the subtle relationships between partial separability
and entanglement of subsystems in multiqubit quantum states and
give experimentally accessible conditions that distinguish between
various classes and levels of partial separability in a
hierarchical order. These conditions
take the form of bounds on the correlations of locally orthogonal
observables. Violations of such inequalities give strong
sufficient criteria for various forms of partial  inseparability and
multiqubit entanglement. The strength of these criteria is
illustrated by showing that they are stronger than several other
well-known  entanglement criteria (the
fidelity criterion, violation of Mermin-type separability inequalities, the
Laskowski-\.Zukowski criterion and the D\"ur-Cirac criterion), and
also by showing their great noise robustness for a variety of
multiqubit states, including $N$-qubit GHZ states and Dicke
states. Furthermore, for $N\geq 3$ they can detect bound entangled
states.  For all these states, the required number of measurement
settings for implementation of the entanglement criteria is shown
to be only $N+1$.  If one chooses the familiar Pauli matrices as
single-qubit observables, the inequalities take  the form of
bounds on the anti-diagonal matrix elements of a state in terms of
its diagonal matrix elements.
\end{abstract}

\author{Michael Seevinck}
\email{m.p.seevinck@uu.nl}
\author{Jos Uffink}
\email{j.b.m.uffink@uu.nl}
\affiliation{Institute of History and Foundations
of Science, Utrecht University,
P.O Box 80.010, 3508 TA Utrecht, The Netherlands}


\maketitle

\section{Introduction}
The problem of characterizing entanglement for multipartite
quantum systems has recently drawn much attention. An important
issue in this problem is that, apart from the extreme cases of
full separability and full entanglement of all particles in the
system, one also has to face the intermediate cases
 in which only some particles in the system are
entangled and others not.  The latter states are usually called
`partially separable' or, more precisely, `$k$-separable' when
they take the form of a  mixture
 of states  that factorize when the
$N$-partite system is partitioned into $k$ subsystems ($k \leq
N$) \cite{duer2,duer,bound3,nagataPRL} .  In this paper we will focus on multiqubit systems only.
 We propose a classification of  partially separable states
for such systems,  slightly extending the classification
introduced by D\"ur and Cirac~\cite{duer2}. This classification
consists of  a hierarchy of levels corresponding to the
$k$-separable states for $k=1, \ldots N$, and within each level
various classes are distinguished by specifying  under which
partitions of the system the state is separable or not.

Several experimentally accessible conditions to characterize
$k$-separable multiqubit states have already been proposed, e.g.,
by Laskowski and \.Zukowski  \cite{laskowzukow}, Mermin-type separability 
inequalities  \cite{nagataPRL,roy, uffink,seevsvet,collins,gisin} or in terms of entanglement witnesses \cite{tothguhne2}. However,
these conditions  do not distinguish the various classes within
the levels.   Separability conditions that do distinguish some of
these classes in the hierarchy  were developed by D\"ur and Cirac.
 Here we present separability conditions that extends and strengthens all the
 conditions just mentioned.

These new conditions take the form of sets of inequalities that bound the
correlations for standard Bell-type experiments (involving at
each site measurement of two orthogonal spin observables). They form a
hierarchy with  bounds that decrease by a factor of four for each
level $k$ in the partial separability hierarchy. For the classes
within a given level, the inequalities give state-dependent
bounds, differing for each class. Violations of the inequalities
provide strong sufficient criteria for various forms of
inseparability and multiqubit entanglement.

We  demonstrate the strength of these conditions in two ways:
Firstly, by showing that they imply several other general
separability conditions, namely the fidelity criterion
\cite{sackett, seevuff,fidelity} \forget{,(also called the
projector based witness with respect to the generalized $N$-qubit
GHZ states)}, the partial separability conditions just mentioned,
i.e.  the Laskowski-\.Zukowski
condition %\cite{laskowzukow}
(with a strict improvement
for $k=2,N$), the D\"ur-Cirac condition %\cite{duer2},
 and the Mermin-type separability
inequalities. We also show that the latter are equivalent to the
Laskowski-\.Zukowski condition.

Secondly, we  compare the conditions to other state-specific
multiqubit entanglement criteria \cite{tothguhne2,guhne2007,chen}
both for their white noise robustness and for the number of
measurement settings required in their implementation. In
particular, we show (i) detection of bound entanglement for $N\geq
3$ with noise robustness for detecting the bound entangled states
of Ref. \cite{bound3} that goes to $1$ for large $N$ (i.e.,
maximal noise robustness), (ii) detection of the
 four qubit Dicke state with noise robustness $0.84$ and $0.36$ for detecting
 it as entangled and fully entangled respectively, (iii) great noise and decoherence robustness
\cite{noise2,noise} in detecting entanglement of the $N$-qubit GHZ
state where for colored noise and for decoherence due to dephasing
the robustness for detecting full entanglement goes to $1$ for
large $N$, and lastly, (iv) better white noise robustness than the
stabilizer witness criteria of Ref. \cite{tothguhne2} for
detecting the $N$-qubit GHZ states. In all these cases it is shown
that only $N+1$ settings are needed.

Choosing the familiar Pauli matrices as the local orthogonal
observables  yields  a convenient matrix  element representation
of the partial separability conditions. In  this representation,
the inequalities give specific bounds on the anti-diagonal matrix
elements in terms of the diagonal ones. Further, some comments
will be made along the way on how these results relate to the
original purpose \cite{bell64} of Bell-type inequalities to test
local hidden-variable models (LHV) models against  quantum
mechanics. Most notably,  when the number of parties is increased,
there is not only
 an exponentially increasing factor that separates the correlations
 allowed in maximally entangled states in comparison to those of local
 hidden-variable theories,
 but, surprisingly, also an exponentially increasing factor between the
 correlations allowed by LHV models and those allowed by
 non-entangled qubit states.

This paper is structured as follows. In section \ref{partsep} we
define the relevant partial separability notions and extend the
hierarchic partial separability classification of Ref.\
\cite{duer2}. There we also introduce the notions of $k$-separable
entanglement  and of $m$-partite entanglement in order to
investigate the relation between partial separability and
multipartite entanglement. We then discuss four known partial
separability conditions discussed above.
 In section  \ref{criteria}  we derive new partial separability
 conditions for $N$ qubits in terms of locally orthogonal
 observables.  They provide the desired
necessary conditions for the full hierarchic separability
classification. In section \ref{strengthsection} the experimental
strength of these criteria is discussed. We end in section
\ref{discussion} with a discussion of the results obtained.

\section{Partial separability and multipartite entanglement}\label{partsep}
In this section we introduce terminology and definitions
to be used in later sections. We define the notions of
$k$-separability, $\alpha_k$-separability, $k$-separable entanglement and $m$-partite
entanglement and use these notions to capture aspects
of the separability and entanglement structure in multipartite
states.   We review  the separability hierarchy
introduced  by D\"ur and
Cirac \cite{duer2} and  extend their classification. We also
discuss four 
partial separability conditions known in the literature \forget{and whose violations are sufficient
criteria for forms of entanglement.}
 These conditions will be strengthened in section III.

\subsection{Partial separability and the separability hierarchy}\label{generalksep}
 Consider an $N$-qubit system  with Hilbert space
 $\H=\mathbb{C}^{2}\otimes\ldots\otimes\mathbb{C}^{2}$.  Let
$ \alpha_k= (S_1, \ldots, S_k)$ denote a partition of $\{ 1,
\ldots ,N\}$ into $k$ disjoint nonempty subsets
  \mbox{($k\leq N$)}. Such a partition  corresponds to a division of
the system into $k$ distinct subsystems, also called a $k$-partite
split \cite{duer2}. A quantum state $\rho$ of this $N$-qubit
system is $k$-separable under a specific $k$-partite split
$\alpha_k$ \cite{duer2,duer,bound3,nagataPRL}  iff it is fully separable in terms of the $k$ subsystems in
this split, i.e., iff
  \beq\label{kseprel}
\rho=\sum_i p_i
  \otimes_{n=1}^{k} \rho_i^{S_n } \,,~~~~~p_i\geq 0,~~\sum_i
p_i =1, \eeq where
 $\rho^{S_n }$ is  a state of subsystem  corresponding to $S_n$ in the split $\alpha_k$.
We denote such states as  $\rho \in {\cal D}_N^{\alpha_k}$ and
also call them $\alpha_k$-separable, for short.  Clearly, ${\cal
D}_N^{\alpha_k}$  is a convex set. A state of the $N$-qubit
system outside this set is
 called $\alpha_k$-inseparable.

 More generally, a state  $\rho$ is called
$k$-separable \cite{laskowzukow,guhnetothbriegel,tothguhne1,acin, haffner} (denoted
as $\rho \in {\cal D}_N^{k\textrm{-sep}}$)
iff there exists a convex decomposition \beq\label{ksep}
\rho=\sum_j p_j
  \otimes_{n=1}^{k}\rho^{S^{(j)}_n } \,,~~~~~p_j\geq 0,~~\sum_j
p_j =1, \eeq where each state $\otimes_{n=1}^{k}\rho^{S^{(j)}_n }$
 is a tensor product of $k$
density matrices of the subsystems corresponding to some such
partition $\alpha^{(j)}_k$, i.e., it factorizes under this split
$\alpha_k^{(j)}$. In this definition, the partition may vary for
each $j$, as long as it is a $k$-partite split, i.e., contains $k$
disjoint non-empty sets. Clearly ${\cal D}_N^{k\textrm{-sep}}$ is
also convex; it is the convex hull of the union  of all $\ {\cal
D}_N^{\alpha_k}$ for fixed values of $k$ and $N$.   States that
are not $k$-separable will be called $k$-inseparable. Note that  a
$k$-separable state need not be $\alpha_k$-separable for any
particular split $\alpha_k$\forget{For example,  consider the
    following two three-qubit states with the three qubits denoted by
    $a$, $b$ and $c$: $\ket{\psi_1}=\ket{0}\otimes(\ket{01}
    -\ket{10})/\sqrt{2}$, $\ket{\psi_2}=\ket{01}
    -\ket{10})/\sqrt{2}\otimes\ket{0}$. The state $\ket{\psi_1}$ is
    biseparable under split $a$-$(bc)$, i.e., $\ket{\psi_1}\in {\cal
    D}_3^{a\textrm{-}(bc)}$, and $\ket{\psi_2}$ is biseparable under
    split $c$-$(ab)$, i.e., $\ket{\psi_2}\in {\cal
    D}_3^{c\textrm{-}(ab)}$. Now form a convex mixture of these two
    states: $\rho=\frac{1}{2}(\ket{\psi_1}\bra{\psi_1}
+\ket{\psi_2}\bra{\psi_2})$. This state $\rho$ is not biseparable
under any split, yet it is by construction biseparable, i.e,
$\rho\in {\cal D}_3^{2\textrm{-sep}}$, and is thus not fully
inseparable \footnote{ Furthermore, using an eigenvalue analysis
of the partial transposed state under each bipartite split, we
obtain that $\rho$ has negative partial transposition (NPT) under
all bipartite splits.  Although NPT under a split is sufficient
for inseparability under this split, NPT under all bipartite
splits is  not sufficient for full inseparability. This is in some
sense analogous to the fact that positive partial transposition
(PPT) with respect all bipartite splits is not sufficient for
full separability.}.} \cite{voet}.
     And even the converse  implication need not hold:
  If a state is biseparable under every bipartition, it does not have
  to be fully separable, as shown
  by the three-partite examples in
 Ref.\ \cite{state}. Similar observations (using different terminology) were presented in Refs. \cite{guhnetothbriegel,tothguhne1}, but below we
 will present a more systematic investigation.

The notion of $k$-separability naturally induces a hierarchic
ordering of the $N$-qubit states. Indeed, the sequence of sets
${\cal D}_N^{k\textrm{-sep}}$ is nested:
 ${\cal D}_N^{N\textrm{-sep}} \subset {\cal D}_N^{(N-1)\textrm{-sep}}
 \subset \cdots \subset
  {\cal D}_N^{1\textrm{-sep}}$.  In other words, $k$-separability implies $\ell$-separability for all $\ell \leq k$.
  We call a  $k$-separable state that is not $(k +1)$-separable  ``$k$-separable entangled''. Thus,  each $N$-qubit state can be characterized by the level  $k$ for which it is $k$-separable entangled, and these levels provide a hierarchical ranking: at one extreme end are the $1$-separable entangled states which are fully  entangled (e.g., the GHZ states), at the other end are the $N$-separable or fully separable states (e.g. product states or   the ``white noise state''  $\openone/ 2^N$).

  Often, it is interesting to know  how many qubits
 are entangled in a $k$-separable entangled state. However, this question does not have a
unique answer.
For example, take $N=4$ and  \mbox{$k=2$} (biseparability). In
this case two types of states may occur in the decomposition
(\ref{ksep}), namely $\rho^{\{ij\}}\otimes\rho^{\{kl\}}$ and
$\rho^{\{i\}}\otimes\rho^{\{jkl\}}$ ($i,j,k,l=1,2,3,4$). A
$2$-separable entangled four-partite state might thus be two- or three-partite
entangled.

In general, an $N$-qubit state $\rho$ will be called $m$-partite
entangled iff  a decomposition of the state such as in
(\ref{ksep}) exists such that each subset $S^{(i)}$ contains at
most $m$ parties, but no such decomposition is possible when all
the $k$ subsets are required to contain less than $m$ parties
 \cite{seevuff}. (In Ref.\ \cite{guhnetothbriegel,tothguhne1} this is called
 `not producible by $(m-1)$-partite entanglement').
 It follows that a $k$-separable entangled state
is also $m$-partite entangled, with $\integer{N/k}\leq m\leq N-k+1$.
Here $\integer{N/k}$ denotes the smallest integer which is not less
than $N/k$. Thus, a state that is
$k$-separably entangled ($k< N$) is at least
$\integer{N/k}$-partite entangled and might be up to
\mbox{$(N-k+1)$}-partite entangled.
Therefore,
conditions that
 distinguish $k$-separability from
\mbox{$(k+1)$}-separability also provide conditions for
 $m$-partite entanglement, but generally allowing a wide range of values of $m$.
For example, for  $N=100$ and $k=2$, $m$ might lie anywhere between 50 and 99.

Of course, a much tighter conclusion about $m$-partite
entanglement can be drawn if we know exactly under which splits
the state is separable. This is why the notion of
$\alpha_k$-separability is helpful, since it provides these finer
distinctions. For example, suppose that a 100-qubit state is
separable under the bipartite split $(\{1\}, \{ 2, \ldots 100\})$
but under no other bipartite split. This state would then  be
$2$-separable (biseparable) but now we could also infer that
$m=99$. On the other hand, if the state were only separable under
the split $(\{ 1, \ldots 50\}, \{51, \ldots 100 \}$, it would still
be biseparable, but only $m$-partite entangled for  $m= 50$.

D\"ur and Cirac \cite{duer2} provided such a fine-grained
classification of $N$-qubit states  by considering their
   separability or inseparability under  all $k$-partite splits.
Let us introduce this classification (with a slight extension) by means of the example of
three qubits,  labeled as $a,b,c$.

\emph{Class 3.} Starting with the lowest level $k=3$, there is
only one 3-partite split, $a$-$b$-$c$,  and consequently only  one
class to be distinguished at this level , i.e. ${\cal
D}_3^{a\textrm{-}b\textrm{-}c}$. This set coincides with ${\cal
D}^{3\textrm{-sep}}_3$.

\emph{Classes 2.1---2.8} Next, at level $k=2$, there are three
bipartite splits: $a$-$(bc)$, $b$-$(ac)$ and $c$-$(ab)$ which
define the sets ${\cal D}_3^{a\textrm{-}(bc)}$, ${\cal
D}_3^{b\textrm{-}(ac)}$, and ${\cal D}_3^{c\textrm{-}(ab)}$.  One
can further distinguish classes defined by all logical combinations of
separability and inseparability under these  splits, i.e. all the
set-theoretical intersections and complements shown in Figure 1.
This leads to classes 2.2 -- 2.8. D\"ur and Cirac showed that all
these classes are non-empty.  To these, we add one more class
2.1: the set of biseparable states that are not separable under
any split. As we have seen, this set is non-empty too.

\emph{Class 1.} Finally,  at level $k=1$ there is again only one
(trivial) split $(abc)$,
 and thus only one class, consisting of all the fully entangled states,  i.e., ${\cal
D}_3^{1\textrm{-sep}} \setminus {\cal D}_3^{2\textrm{-sep}}$.

\begin{figure}[h]
\includegraphics[scale=0.45]{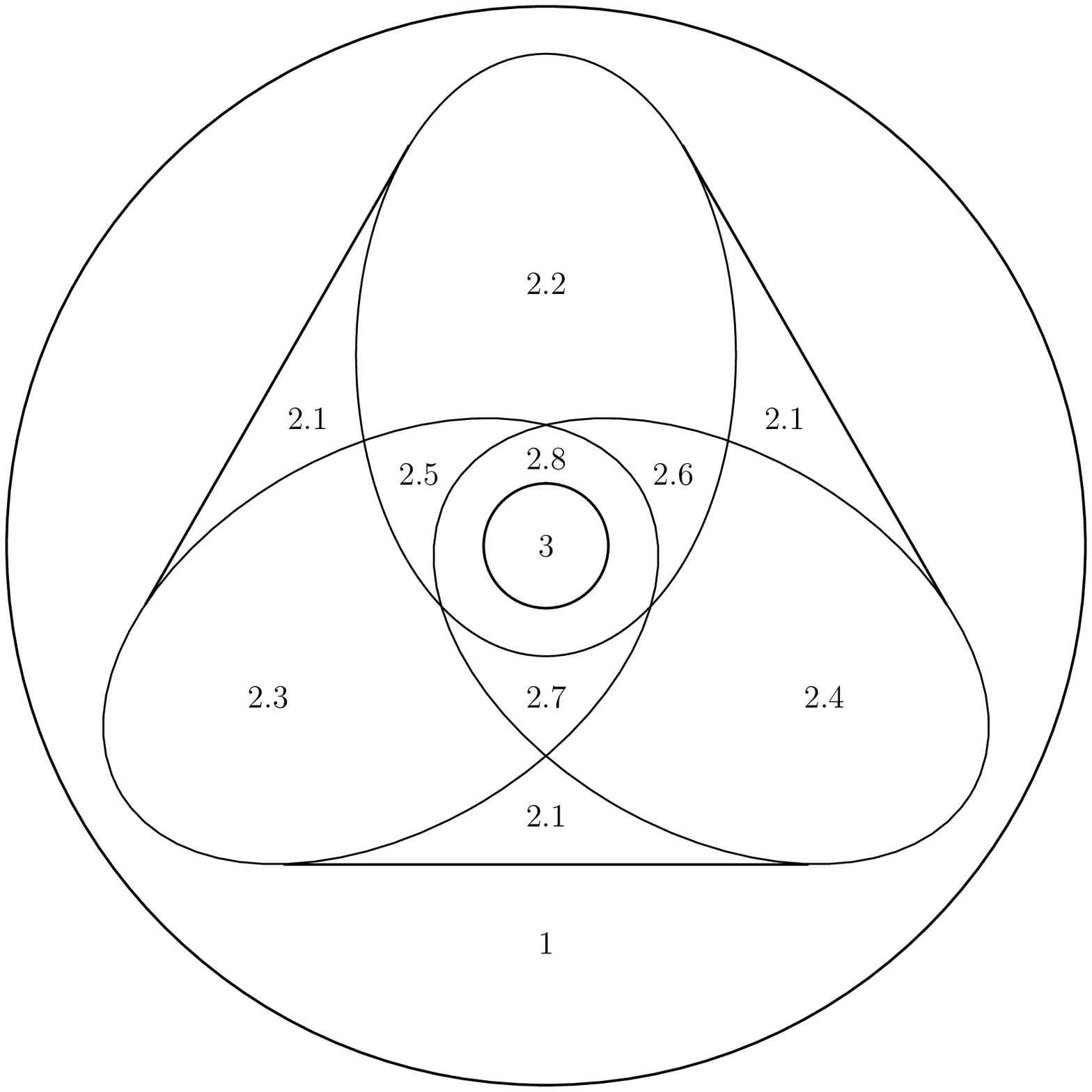}
\caption{Schematic representation of the 10 partial separability
classes of three-qubit states} \label{figclass}
\end{figure}

We feel that the above extension is desirable since otherwise  the
D\"ur-Cirac classification would not
 distinguish between class 2.1 and class 1. However,  states in class 2.1 are
simply convex combinations of states that are biseparable under
different bipartite splits. Such states can be realized by mixing
the biseparable states, and are conceptually different from the
fully inseparable states of class 1.

This three-partite example serves to illustrate how the D\"ur-Cirac separability
classification works for general $N$. Level $k$ ($1\leq k \leq N$)
of the separability hierarchy consists of all $k$-separable
entangled states.  Each level is further divided into distinct
classes by considering all logically possible combinations of
separability and inseparability under the various $k$-partite
splits. The number of such classes increases rapidly with $N$, and
therefore we will not attempt to list them. In general, all such
classes may be non-empty.   As an extension of the D\"ur-Cirac
classification, we distinguish at each level $1< k<N$ one further
class, consisting of $k$-separable entangled states that are not
separable under any $k$-partite split.

  \forget{
A different but equivalent formulation of this feature uses the
notion of semi-separability \cite{reviewHHHH} (semi-separable
states are separable under all splits $\{ 1 \}\{ 2,\ldots,N-1\}$):
there exist semi-separable states that are still entangled. In
general, $k$-separability implies $(k-1)$-separability but  even
$(k-1)$-separability under all $(k-1)$-partite splits is not
 sufficient for $k$-separability \emph{simpliciter}, and neither is
 $k$-inseparability under all
$k$-partite splits sufficient for $(k-1)$-inseparability
\emph{simpliciter}.}

\forget{ At each level $k$ of the hierarchy various classes exist,
which are obtained by all possible combinations of
$k$-separability and $k$-inseparability under all $k$-partite
splits, as well as by considering the possibility of
$k$-separability \emph{simpliciter}.  States belong to the same
class if they are (in-)separable with respect to the same
$k$-partite splits. However, one class on this level is not
captured in this way, namely the class of states that are
inseparable under all $k$-partite splits at this level yet that
are nevertheless $k$-separable \emph{simpliciter} and not
$(k+1)$-separable \emph{simpliciter}.  As mentioned before, this
latter class is not obtained in the D\"ur-Cirac classification.
Compared to their classification, at each level such an extra
class is obtained. For example, four-partite states that are
biseparable only under a single bipartite split (e.g,
$(ab)$-$(bc)$ or $a$-$(bcd)$) are each in different classes on
level $2$, and states that are inseparable under all these
bipartite splits but that are nevertheless $2$-separable
\emph{simpliciter} but not $3$-separable \emph{simpliciter} also
belong to a different class on level $2$. }

In order to find relations between these classes, the notion of a
\emph{contained split} is useful \cite{duer2}. A $k$-partite split
$\alpha_k$ is contained in a $l$-partite split $\alpha_l$, denoted
as $\alpha_k \prec \alpha_\ell$,  if $\alpha_l$  can be obtained
from $\alpha_k$ by joining some of the subsets of $\alpha_k$. The
relation $\prec$ defines a partial order between splits at
different levels. \forget{On the one hand, a number of different
$k$-partite splits may be contained in the same $l$-partite split,
and on the other hand, each $k$-partite split is contained in
various $l$-partite splits (See further Ref. \cite{duer2}).} This
partial order is  helpful because $\alpha_k$-separability implies
$\alpha_\ell$-separability of all splits $\alpha_\ell$ containing
$\alpha_k$. \forget{and $\alpha_\ell$-inseparability implies
inseparability under all splits that are contained in $\alpha_l$.}
We will use this implication below to obtain conditions for
separability of a $k$-partite split at level $k$ from such
conditions on all  $(k-1)$-partite splits at level $k-1$ this $k$-partite
split is contained in. \forget{One can thus construct separability
conditions for all classes at higher levels from the separability
conditions for classes at level $k=2$.}

\forget{
A an example of this  hierarchic  separability classification
consider the case of three qubits. The classification under
specific splits is given in Ref.\ \cite{duer2}, but we extend it
here to also include the classification under $k$-separability
\emph{simpliciter}. In the next section we give separability
conditions for this full classification. To perform the
classification one  considers all possible $3$-partite splits of
the three qubits, say $a,b,c$. We denote the three-partite split
where each qubit is contained in a separate subset as $a$-$b$-$c$,
and bipartite splits analogously as $a$-$(bc)$, $b$-$(ac)$ and
$c$-$(ab)$. We next need to consider the separability properties
of a state $\rho$ under each of these splits as well as the
separability \emph{simpliciter} properties. We start with the
lowest level, level $k=3$. Here one determines whether the state
$\rho$ is separable under the $3$-partite split $a$-$b$-$c$. If
indeed so this state is $3$-separable under this split, and can be
written as $\rho= \sum_j p_{j}\rho_{j}^{(a)}\otimes\rho_j^{(b)}
\otimes\rho_{j}^{(c)}$ (with $\sum_jp_j=1$, $0\leq p_j\leq 1$). On
the next level, level $k=2$, separability under the bipartite
splits is firstly investigated. Thus, a state is biseparable
under the split $a$-$(bc)$ if it can be written as $\rho= \sum_j
p_{j}\rho_{j}^{(a)}\otimes\rho_{j}^{(bc)}$, and analogous for the
other two bipartite splits.  Each combination of separability or
inseparability for these various
 bipartite splits corresponds to a class in the classification.
We are left with classifying states that are not separable under
any bipartite split. D\"ur and Cirac have put these in a single
class and called them fully inseparable since they can not be
written in terms of definition (\ref{kseprel}) for any $k$.
However, such states can still be biseparable \emph{simpliciter},
although not $3$-separable \emph{simpliciter}. Indeed, we have
seen that such states exist and they together make up a disjoint
class as well.  Finally, if a state is also not biseparable
\emph{simpliciter} one obtains the disjoint class that consists of
states that are fully inseparable, i.e., they cannot be written in
terms of definition (\ref{ksep}) for any $k$.  In this way a total
of $10$ disjoint classes can be found. This classification is
denoted as follows:} \forget{
\begin{table}[!h]
\begin{tabular}{|l|l|l|} \hline
Class &  Separability conditions \\ \hline 1 &  Fully inseparable
states. & \parbox{6cm}{
 States that are not $k$-separable (for $k\geq2$), i.e., that  cannot be written in the form (\ref{ksep}),
 i.e., $\rho \notin {\cal D}_3^{2\textrm{-sep}}$. An example is the GHZ state $(\ket{000} \pm\ket{111})/\sqrt{2}$.} \\
2.1&    States that are biseparable \emph{simpliciter}.&
\parbox{6cm}{ can thus be written in the form (\ref{ksep}) for $k=2$,
but not for $k=3$, and that are not separable under any bipartite
split, i.e., they cannot be written in the form (\ref{kseprel})
for any $k>1$, i.e, $\rho \in {\cal D}_3^{2\textrm{-sep}}$, $\rho
\notin {\cal D}_3^{a\textrm{-}(bc)} \cup{\cal
D}_3^{b\textrm{-}(ac)} \cup{\cal D}_3^{c\textrm{-}(ab)}\cup{\cal
D}_3^{3\textrm{-sep}}$. As we have already seen, this class is not
empty.}
                \\
 2.2 &   1-qubit biseparable states.  &
\parbox{6cm}{States in class $2.2$ are separable under split $a$-$(bc)$, but
inseparable under the other two bipartite splits, i.e., $\rho \in
{\cal D}_3^{a\textrm{-}(bc)}, \rho \notin{\cal
D}_3^{b\textrm{-}(ac)}\cup{\cal D}_3^{c\textrm{-}(ab)}$.
Similarly, states in classes $2.3$ and $2.4$  are separable under
split $b$-$(ac)$ and split $c$-$(ab)$ respectively, and
inseparable otherwise.}     \\
 2.3 &  &  \\
 2.4 &  & \\
 2.5  &
 2-qubit biseparable states.& \parbox{6cm}{
States in class $2.5$  are biseparable under split  $a$-$(bc)$
and $b$-$(ac)$ but inseparable under split $c$-$(ab)$, i.e., $\rho
\in {\cal D}_3^{a\textrm{-}(bc)} \cup{\cal D}_3^{b\textrm{-}(ac)},
\rho\notin{\cal D}_3^{c\textrm{-}(ab)}$ Similarly, states in
classes $2.6$  ($2.7$)  are biseparable under splits $a$-$(bc)$
and  $c$-$(ab)$ ($b$-$(ac)$ and $c$-$(ab)$), but inseparable
otherwise }   \\
2.6  &  &
 \\
2.7 &  & \\
2.8 &  3-qubit biseparable states.  &\parbox{6cm}{States that are
separable under all bipartite splits, but that are inseparable
under the three-partite split $a$-$b$-$c$, i.e., $\rho \in {\cal
D}_3^{a\textrm{-}(bc)} \cup{\cal D}_3^{b\textrm{-}(ac)} \cup{\cal
D}_3^{c\textrm{-}(ab)}, \rho \notin  {\cal D}^{3\textrm{-sep}}_3$.
This class is not empty, as shown by the examples of Refs.
\cite{state}}.
\\
3 & fully separable states. &\parbox{6cm}{States that are
separable under all splits, i.e., $\rho \in  {\cal
D}^{3\textrm{-sep}}_3$. An example is the state $\ket{000}$. That
classes $2.2-2.7$ are not empty was shown in \cite{duer2}.}\\
 \hline
\end{tabular}
\caption{Separability classes in the full separability
classification of three qubit states.} \label{table0}
\end{table} }

\forget{ \emph{Class $1$:  Fully inseparable states} ($\rho \notin
{\cal D}_3^{2\textrm{-sep}}$).
 States  that  cannot be written in the form (\ref{ksep}) for $k\geq 2$.
  An example is the GHZ state $(\ket{000} \pm\ket{111})/\sqrt{2}$.

\emph{Class $2.1$: biseparable states that are inseparable under
all bipartite splits}
 $\rho \in {\cal
D}_3^{2\textrm{-sep}}$, $\rho \notin {\cal D}_3^{a\textrm{-}(bc)}
\cup{\cal D}_3^{b\textrm{-}(ac)} \cup{\cal
D}_3^{c\textrm{-}(ab)}\cup{\cal D}_3^{3\textrm{-sep}}$. States
that can  be written in the form (\ref{ksep}) for $k=2$, but not
for $k=3$, and that are not separable under any bipartite split,
i.e., they cannot be written in the form (\ref{kseprel}) for any
$k>1$.  As we have seen, this class is not empty.

\emph{Classes $2.2$, $2.3$, $2.4$: 1-qubit biseparable states}.
States in class $2.2$ are separable under split $a$-$(bc)$, but
inseparable under the other two bipartite splits, i.e., $\rho \in
{\cal D}_3^{a\textrm{-}(bc)}, \rho \notin{\cal
D}_3^{b\textrm{-}(ac)}\cup{\cal D}_3^{c\textrm{-}(ab)}$.
Similarly, states in classes $2.3$ and $2.4$  are separable under
split $b$-$(ac)$ and split $c$-$(ab)$ respectively, and
inseparable otherwise.

\emph{Classes $2.5$, $2.6$, $2.7$: 2-qubit biseparable states}.
States in class $2.5$  are biseparable under split  $a$-$(bc)$
and $b$-$(ac)$ but inseparable under split $c$-$(ab)$, i.e., $\rho
\in {\cal D}_3^{a\textrm{-}(bc)} \cup{\cal D}_3^{b\textrm{-}(ac)},
\rho\notin{\cal D}_3^{c\textrm{-}(ab)}$ Similarly, states in
classes $2.6$  ($2.7$)  are biseparable under splits $a$-$(bc)$
and  $c$-$(ab)$ ($b$-$(ac)$ and $c$-$(ab)$), but inseparable
otherwise.

\emph{Class $2.8$: 3-qubit biseparable states}. States that are
separable under all bipartite splits, but that are inseparable
under the three-partite split $a$-$b$-$c$, i.e.,  $\rho \in {\cal
D}_3^{a\textrm{-}(bc)} \cup{\cal D}_3^{b\textrm{-}(ac)} \cup{\cal
D}_3^{c\textrm{-}(ab)}, \rho \notin  {\cal D}^{3\textrm{-sep}}_3$.
This class is not empty, as shown by the examples of Refs.
\cite{state}.

\emph{Class  $3$: fully separable states}. States that are
separable under all splits, i.e., $\rho \in  {\cal
D}^{3\textrm{-sep}}_3$. An example is the state $\ket{000}$.
\noindent That classes $2.2-2.7$ are not empty is shown in
\cite{duer2}.

Here we have not followed the notation of D\"ur-Cirac \cite{duer2}
since they do not distinguish between class $1$ and $2.1$. We
believe that our notation  better reflects the three different
levels.
 Note that  states in the classes  $2.1$ to $4$ are in ${\cal
D}^{2\textrm{-sep}}_3$, whereas states in class $3$  are in the
set ${\cal D}^{3\textrm{-sep}}_3$ and states in class $1$ are in
neither of these two separability sets.}

\forget{
\subsection{$k$-separability and $m$-partite entanglement in $N$-partite systems}

The $k$-separability properties alone do not determine the
multipartite entanglement properties completely, but some
interesting relations exist nevertheless, as we will outline
below.

For example, if we look at the classification of three qubits given
above, we see that states in class $1$ are $1$-partite entangled,
also called fully entangled, states in classes $2.1$ to $2.8$ are $2$-partite entangled,  and states in
class $3$ are fully unentangled.

} The multi-partite entanglement properties of  $k$-separable  or
$\alpha_k$-separable  states are subtle, as can be seen from the
following examples.

(i) mixing states does not conserve $m$-partite entanglement. Take
$N=3$, then mixing the $2$-partite entangled $2$-separable states
$\ket{0}\otimes(\ket{00}+\ket{11})/\sqrt{2}$ and
$\ket{0}\otimes(\ket{00} -\ket{11})/\sqrt{2}$ with equal weights
gives  a $3$-separable state $(\ket{000}\bra{000}
+\ket{011}\bra{011})/2$.

(ii) an $N$-partite state can be $m$-partite entangled ($m<N$)
even if it has no $m$-partite subsystem whose (reduced) state is
$m$-partite entangled \cite{seevuff,guhnetothbriegel}. Such states are said to have
irreducible $m$-partite entanglement \cite{walck}. Thus, a state
of which some reduced state is $m$-partite entangled is itself at
least $m$-partite entangled,
 but the converse need not be true.

(iii)  consider a biseparable entangled state that is  only 
separable under the bipartite split $(\{1\},\{2,\ldots,N\})$. One cannot infer
that the subsystem $\{2,\ldots,N\}$ is \mbox{$(N-1)$}-partite
entangled. A counterexample is
 the three-qubit state $\rho = (
\ket{0}\bra{0}\otimes P^{(bc)}_{-} + \ket{1}\bra{1}\otimes
P^{(bc)}_{+})/2
 $
which is biseparable only under the partition $a\textrm{-}(bc)$,
and thus bipartite entangled, but has no bipartite subsystem
whose reduced state is entangled. Here $P^{(bc)}_{+}$ and
$P^{(bc)}_{-}$ denote projectors on the Bell states
$\ket{\psi^{\pm}}=\frac{1}{\sqrt{2}}(\ket{01} \pm \ket{10})$ for
parties $b$ and $ c$, respectively.

(iv)  a state that is inseparable under all splits but which is
not fully inseparable (i.e., $\rho \in  {\cal
D}^{k\textrm{-sep}}_N$ with $k>1$ and  $\rho\notin\cup_{\alpha_k}
{\cal D}^{\alpha_k}_N$, $\forall \,\alpha_k,k$) might still have all forms of $m$-partite
entanglement apart from full entanglement, i.e., it could be
$m$-partite entangled with $2\leq m\leq N-1$. Thus the state could
even have $m$-partite entanglement as low as $2$-partite
entanglement, although it is inseparable under all splits.  For
example, T\'oth and G\"uhne \cite{tothguhne1} consider a mixture
of two $N$-partite states where each of them is $(\lceil N/2
\rfloor)$- separable according to different splits. This mixed
state is by construction $(\lceil N/2 \rfloor)$- separable, not
biseparable under any split, yet only $2$-partite entangled. See
also  the example in footnote~\cite{voet} which is $(N-1)$-separable and only $2$-partite entangled.

(v) Lastly, $N$-partite fully entangled states exist where no
$m$-partite reduced state is entangled (such as
 $N$-qubit GHZ state) and also where all $m$-partite reduced states
 are entangled (such as the $N$-qubit W-states) \cite{durR}.

These  examples serve to emphasize that one should be very
cautious in inferring the existence of entanglement in subsystems
of a larger system which is known to be $m$-partite entangled or $k$-separable entangled for some specific value of $m$ and $k$.

\subsection{Separability Conditions}\label{introsepconditions}
We now review four separability conditions for qubits,  which will
all be strengthened in the next section. These are necessary
conditions for  states  to be $k$-separable, $2$-separable, and
$\alpha_k$-separable respectively.

(I) Laskowski and \.Zukowski \cite{laskowzukow} showed that for
any $k$-separable $N$-qubit state  $\rho$ the anti-diagonal matrix
elements (denoted by $\rho_{j,\bar{\jmath}}$, where
$\bar{\jmath}=d+1-j$, $d =2^N$ ) must satisfy
 \beq\label{antidiagonal}  \max_j | \rho_{j,\bar{
 \jmath}}|  \forget{, \rho_{2, (d-1)},
 \ldots , \rho_{d, 1}\}
|\rho_{\nearrow}^{}|}
\leq
\big(\frac{1}{2}\big{)}^{k},~~~\forall\rho_{} \in {\cal
D}^{k\textrm{-sep}}_N. \eeq
 This condition can be
 easily proven by the observation that
for any density matrix to be physically meaningful its
anti-diagonal matrix elements must not exceed $1/2$ . Therefore,
anti-diagonal elements of a product of $k$ density matrices cannot
be greater than $(1/2)^{k}$. \forget{For pure states this can be
checked directly \forget{\cite{footnoteproof}}
 and then, } By convexity, this results then holds
all $k$-separable states. Note that this condition is not basis
dependent.

 It follows  from (\ref{antidiagonal}) that if the anti-diagonal matrix elements
of state $\rho$ obey
 \beq\label{antidiagonalentang}
 \big(\frac{1}{2}\big)^{k} \geq  \max_j
 |\rho_{j,\bar{\jmath}}|>   \big(\frac{1}{2}\big)^{k+1},
\eeq then  $\rho$ is at most $k$-separable, i.e.,  $k$-separable
entangled, and thus at least $m$-partite entangled, with
$m\geq\integer{N/k}$.

 \forget{
Thus,
$\max |\rho_{\nearrow}|>{1}/{4}$
 is a sufficient condition for full $N$-partite entanglement,
  $ {1}/{4} \geq\max |\rho_{\nearrow}|>{1}/{8}$   is sufficient for  $2$-separable entanglement  (and at least $m$-partite
entangled with $m\geq\integer{N/2}$), and so on. If furthermore
the number of parties per subset in the partition is specified,
i.e., if a specific $k$-partite split is known under which the
state is separable, we get sufficient $m$-partite entanglement
criteria for a \emph{definite} number $m$. To choose an example,
suppose a state is biseparable under the bipartition of the form
$\{1\}\{2,\ldots,N\}$. Then the state is \mbox{$(N-1)$}-partite
entangled, whereas if it is separable under a bipartition  that
splits the set of parties in half (i.e., $\{1,\ldots,
\integer{N/2}\}\{\integer{N/2}+1,\ldots ,N\}$) the state is only
$\integer{N/2}$-partite entangled.
}
The partial separability condition (\ref{antidiagonal}) does not
yet explicitly refer to directly experimentally accessible
quantities. However, in the next section we will rewrite this
condition in terms of expectation values of local observables, and
show that they are  equivalent to Mermin-type separability
inequalities. \forget{ We strengthen this condition for $k=2$ and
$k=N$ considerably. This yields stronger sufficient criteria for
when states are entangled or fully entangled respectively. For
other $k$ it is unknown if the new conditions yield stronger
sufficient criteria for $k$-inseparability.}

(II) Mermin-type separability inequalities  \cite{nagataPRL,roy, uffink,seevsvet,collins,gisin}. Consider the familiar CHSH operator for two qubits (labeled as $a$ and $b$) which is defined by: \beq
 M^{(2)} :=
X^{}_a\otimes X^{}_b + X^{}_a\otimes Y^{}_b + Y^{}_a\otimes
X^{}_b- Y^{}_a\otimes Y^{}_b. \label{1} \eeq  Here, $X^{}_a$ and $Y^{}_a$ denote two spin observables on the Hilbert spaces ${\cal H}_a$ and ${\cal  H}_b$ of qubit $a$, and $b$. The so-called
Mermin operator \cite{mermin} is  a generalization of this operator to $N$
qubits (labeled as $(a, b, \ldots n)$), defined by the recursive relation: \beq \label{merminN}
M^{(N)}:= \frac{1}{2}M^{(N-1)} \otimes(X^{}_n +Y^{}_n) +
 \frac{1}{2}M'^{(N-1)}\otimes(X^{}_n -Y^{}_n),
\eeq where $M'$ is the same operator as $M$ but with
all  $X$'s and $Y$'s  interchanged.

In the special case where, for each qubit, the spin observables $X$ and $Y$ are orthogonal, i.e. $ \{ X_i, Y_i\} =0$ for $i\in \{ a, \ldots n\}$,  Nagata et al.~\cite{nagataPRL} obtained the following $k$-separability
 conditions:
 \beq \label{quadraticN} \av{M^{(N)}}^2
+\av{M'^{(N)}}^2
\leq2^{(N+3)}\big(\frac{1}{4}\big)^k,~~ \forall \rho
\in {\cal D}_N^{k\textrm{-sep}}. \eeq
As just mentioned,  the next section will show that  these inequalities are equivalent to the
Laskowski-\.Zukowski inequalities.  The quadratic inequalities  (\ref{quadraticN}) also imply the following sharp linear Mermin-type
inequality  for $k$-separability: \beq\label{linearN}
|\av{M^{(N)}}|\leq 2^{(\frac{N+3}{2})}\big(\frac{1}{2}\big)^k,~~
\forall \rho \in {\cal D}_N^{k\textrm{-sep}}. \eeq \forget{This
inequality  is sharp: $\sup_{\rho \in {\cal D}_N^{k\textrm{-sep}}}
|\av{M_N} | = 2^{(N+3-2k)/2}$. } \forget{If (\ref{linearN}) is
violated the $N$ qubit state $\rho$ is $(k-1)$-separably
 entangled and it has at least $m$-partite entanglement,
 with $m\geq \integer{N/(k-1)}$.} For $k=N$ inequality (\ref{linearN})  reproduces a
 result obtained by Roy \cite{roy}.

(III). The fidelity $F(\rho)$ of a $N$-qubit state $\rho$ with
respect to the generalized $N$-qubit GHZ state $\ket{\Psi_{{\rm
GHZ},\alpha}^N}:= ( \ket{0}^{\otimes N}
+e^{i\alpha}\ket{1}^{\otimes N})/\sqrt{2}$ ($\alpha \in
\mathbb{R}$) is defined as
\begin{equation}
 F(\rho) := \max_\alpha \langle \Psi_{{\rm GHZ},\alpha}^N |\rho|\Psi_{{\rm GHZ},\alpha}^N\rangle=
 \frac{1}{2} (\rho_{1,1}+\rho_{d,d})
 +|\rho_{1,d}|,
 \label{fidelity}
\end{equation} The fidelity
condition \cite{sackett,seevuff,fidelity} (also known as the
projection-based witness \cite{tothguhne2}) says that for all
biseparable $\rho$:
  \beq  F(\rho) \leq 1/2, ~~~ \forall \rho \in {\cal D}_N^{2\textrm{-sep}}. \label{fidelitycriterion}  \enq
In other words,  $ F(\rho) > 1/2 $
  is a sufficient condition for full $N$-partite entanglement.
An equivalent formulation of (\ref{fidelitycriterion})  is:
\beq\label{equivfidelity} 2|\rho_{1,d}| \leq \sum_{j\neq 1, d}
\rho_{j,j},~~~ \forall\rho_{} \in {\cal D}^{2\textrm{-sep}}_N
.\eeq

 Of course, analogous conditions may be obtained by
replacing $\ket{\Psi_{{\rm GHZ},\alpha}^N}$ in the definition
(\ref{fidelity}) by any other maximally entangled state
\cite{fidelity,Nagata2002}.  Exploiting this feature,  one can
reformulate (\ref{equivfidelity}) in a basis-independent form:
\beq\label{equivfidelity2} 2 \max_{j} |\rho_{j,\bar{\jmath}}| \leq
\sum_{i\neq j,\bar{\jmath} } \rho_{i,i},~~~ \forall\rho_{} \in
{\cal D}^{2\textrm{-sep}}_N .\eeq

Note that in contrast to the Laskowski-\.Zukowski condition and
the Mermin-type separability inequalities, the fidelity condition does not
distinguish biseparability and other forms of $k$-separability.
Indeed, a fully separable state (e.g. $\ket{0^{\otimes N}}$ can
already attain the value $F(\rho)=1/2$. Thus, the fidelity
condition only distinguishes full inseparability \mbox{(i.e.,
$k=1$)} from other types of separability ($k\geq 2$). However, as
will be shown in the next section, violation of the fidelity
condition yields a stronger test for full entanglement than
violation of the Laskowski-\.Zukowski condition.

(IV) The D\"ur-Cirac depolarization method \cite{duer,duer2} gives
necessary conditions for partial separability under specific
bipartite splits. It uses a two-step procedure in which a general
state $\rho$ is first depolarized to become a member of a special
family of states, called $\rho_N$, after which this depolarized
state is tested for $\alpha_2$-separability under a  bipartite
split $\alpha_2$. If the depolarized state $\rho_N$ is not
separable under  $\alpha_2$, then neither is the original state
$\rho$, but not necessarily vice versa since the depolarization
process can decrease inseparability.

 The special family of states $\rho_N$ is  given by
\beq\label{DuerStates}
\rho_N= \lambda_0^+\ket{\psi_0^+}\bra{\psi_0^+} +\lambda_0^-\ket{\psi_0^-}\bra{\psi_0^-}+
\sum_{j=1}^{2^{N-1}-1}\lambda_j(\ket{\psi_j^+} \bra{\psi_j^+}+\ket{\psi_j^-}\bra{\psi_j^-}),
\eeq
with the so-called orthonormal GHZ-basis $\ket{\psi_j^{\pm}}=
\frac{1}{\sqrt{2}}\ket{j0} \pm\ket{j'1})$,
  where $ j=j_1j_2\ldots j_{N-1}$ is in binary notation (i.e., a string of
$N-1$ bits),  and $j'$ means a bit-flip of $j$:  $
j'=j'_1j'_2\ldots j'_{N-1}$, with $j'_i=1,0$ if $j_i=0,1$.
\forget{Note that $\ket{\psi_0^+}$ and $\ket{\psi_0^-}$ are the
states $\ket{\Psi_{{\rm GHZ},\alpha}^N}$ with $\alpha =0,\pi$
respectively.} The depolarization process does not  alter the
values  of
$\lambda_0^{\pm}=\bra{\psi_0^{\pm}}\rho\ket{\psi_0^{\pm}}$ and of
$\lambda_j=(\bra{\psi_j^{+}}\rho\ket{\psi_j^{+}}+\bra{\psi_j^{-}}\rho\ket{\psi_j^{-}})/2$
of the original state $\rho$. The values of $j=j_1j_2\ldots
j_{N-1}$ can be used to label the various bipartite splits by
stipulating that $j_n=0,(1)$ corresponds
to the $n$-th qubit belonging (not belonging) to the same subset as
the last qubit. For example, the splits $a$-$(bc)$, $b$-$(ac)$,
$c$-$(ab)$ have labels $j=10,01,11$ respectively.

The D\"ur-Cirac condition \cite{duer2} says that a state $\rho$ is
separable under a specific bipartite split $j$ if
\beq\label{dccondition} |\lambda_0^+-\lambda_0^-|\leq2
\lambda_j~~~\Longleftrightarrow~~~ 2|\rho_{1,d}|\leq \rho_{l,l}
+\rho_{\bar{l},\bar{l}},~~~~~ \forall\rho_{} \in {\cal D}^{j}_N,
~~~ \bar{l} = d+1 -l,  \eeq For the states (\ref{DuerStates}) this
condition is in fact necessary and sufficient. In the right-hand
side of the second inequality of (\ref{dccondition}) $l$ is
determined from $j$ using Tr$[\rho \ket{\psi_j^+}
\bra{\psi_j^+}+\ket{\psi_j^-}\bra{\psi_j^-}]= \rho_{l,l}
+\rho_{\bar{l},\bar{l}}$.

Separability conditions for multipartite splits are  constructed
from the conditions (\ref{dccondition}) by means of the partial
order $\prec$ of containment. As mentioned above, if a state is
$\alpha_k$-separable, then it is also $\alpha_2$-separable for all
bipartite splits  $\alpha_k \prec \alpha_2$. Therefore,  the
conjunction of all $\alpha_2$-separability conditions  must hold
for such a state.

 \forget{ Thus if $
|\lambda_0^+-\lambda_0^-|>2 \lambda_j$ the state is inseparable
under the split $j$.}
 Note that if
$|\lambda_0^+-\lambda_0^-|>2\max_{j} \lambda_j$,  the state is
inseparable under all bipartite splits, but this does not imply
that it is fully inseparable (cf.\ footnote~\cite{voet}).
 Indeed,  this feature also exists for states  of the form (\ref{DuerStates}) as the following example shows.
 Take the following two members of the  family (\ref{DuerStates}) for $N=3$:
for
 $\rho_3^i$ we choose $\lambda_0^+=1/2$, $\lambda_0^- =0$, $\lambda_{01}=0$, $\lambda_{10}=1/4$, $\lambda_{11}=0$,
 and for
 $\rho_3^{ii}$ :  $\lambda_0^+=1/2$, $\lambda_0^- =0$, $\lambda_{01}=0$, $\lambda_{10}=0$, $\lambda_{11}=1/4$.
\forget{From the results in Ref.  \cite{duer2} it follows that a
state $\rho_3$ has PPT and is also separable under the split
$a$-$(bc)$ iff  $2\lambda_{10}\geq |\lambda_0^+ - \lambda_0^-| $,
has PPT and is separable under the split  $b$-$(ac)$ iff
$2\lambda_{01}\geq |\lambda_0^+ - \lambda_0^-| $, has PPT and is
separable under the split  $c$-$(ab)$ iff $2\lambda_{11}\geq
|\lambda_0^+ - \lambda_0^-| $. Furthermore the state is
$3$-separable iff all bipartite splits are separable. Applying
these results we obtain that}  It follows from condition
(\ref{dccondition}) that $\rho_3^i$ is separable under split
$a$-$(bc)$ and inseparable under  other  splits, while
$\rho_3^{ii}$ is separable under the split $c$-$(ab)$ and
inseparable under any other split. Now form a convex mixture of
these two states: $\tilde{\rho}_3 =\alpha\rho_3^i +\beta
\rho_3^{ii}$ with $\alpha +\beta =1$ and $\alpha,~ \beta \in
(0,1)$.  This state $\tilde{\rho}_3$ is still of the form
(\ref{DuerStates})\forget{with values $\lambda_0^+=1/2$,
$\lambda_0^- =0$, $\lambda_{01}=0$, $\lambda_{10}=\alpha/4$,
$\lambda_{11}=\beta/4$}, so that we can again  apply condition
(\ref{dccondition}) to conclude that $\tilde{\rho}_3$ is not
separable under any bipartite split, yet biseparable by
construction.

 In
the next section we give necessary conditions for $k$-separability
and $\alpha_k$-separability   that are stronger than the
Laskowski-\.Zukowski condition (for $k=2,N$), the fidelity
condition and the D\"ur-Cirac condition. \forget{ so that violation of these new
conditions yields stronger sufficient criteria for inseparability
under splits.}

\section{Deriving new partial separability conditions} \label{criteria}

This section presents separability conditions for all levels and
classes in the separability hierarchy of $ N$-qubit states.
 We
start  with the case of $N=2$, which has been treated more
extensively in \cite{uffseev}.   We next move on to the slightly
more complicated case of three qubits, for which explicit
separability conditions are given for each of the 10 classes in
the separability hierarchy which were depicted in Figure \ref{figclass}. Finally, the case of $N$ qubits is
treated by a straightforward  generalization.

\subsection{Two-qubit case: setting the stage}\label{twoqubitsection}
 For  two-qubit systems the separability
hierarchy is very simple: there is only one possible split, and
consequently just one class at each of the two levels $k=1$ and
$k=2$, i.e., states are either inseparable (entangled) or
separable.

Consider a system composed of a pair of qubits  \forget{on the
Hilbert space $\mathcal{H}=\mathbb{C}^2 \otimes \mathbb{C}^2$} in
the familiar setting of two  distant sites, each receiving one of
the two qubits, and where, at each site, a  measurement of either
of two spin observables is made. We will  focus on the special
case that these local spin observables are mutually orthogonal.
 Let $(X^{(1)}_a, Y^{(1)}_a, Z^{(1)}_a)$ denote three orthogonal spin observables on  qubit  $a$,
 and    $(X^{(1)}_b, Y^{(1)}_b, Z^{(1)}_b)$ on qubit $b$.
(The superscript $1$ denotes that we are
 dealing with  single-qubit operators.)
 A familiar choice  for the orthogonal
triples $\{ X^{(1)},Y^{(1)},Z^{(1)} \}$  are the Pauli matrices
$\{\sigma_x,\sigma_y,\sigma_z\}$.  But note that the choice of the two sets need not coincide.
We further define $I^{(1)}_{a,b}:=\1$. For all single-qubit pure states $\ket{\psi}$  we have
\beq\label{spin}\av{X^{(1)}_j}_{\psi}^2+\av{Y^{(1)}_j}_{\psi}^2+
\av{Z^{(1)}_j}_{\psi}^2=\av{I^{(1)}_j}_{\psi}^2, ~~~j=a,b, \eeq and for mixed
states $\rho$ \beq
\label{spin2}\av{X^{(1)}_j}^2+\av{Y^{(1)}_j}^2+\av{Z^{(1)}_j}^2\leq
\av{I^{(1)}_j}^2, ~~~ j=a,b. \eeq

  We write $X_a X_b$  or even $XX$ etc.\, as shorthand for $X_a\otimes X_b$ and
$\av{X X}:= \mathrm{Tr}[\rho X_a\otimes X_b]$ for the
expectation value  in a general  state $\rho$, and $\av{XX}_\Psi := \bra{\Psi}X_a \otimes X_b \ket{\Psi}$  for the
expectation in  a pure state $\ket{\Psi}$.

\forget{ We also assume local orthogonality of the spin
observables
 i.e., $A\perp A'\perp A''$  (where we have included a third orthogonal observable $A''$
 orthogonal to the first two), and $B\perp B'\perp B''$.
\forget{(for the case of two qubits this amount to the local
observables anticommuting with each other: $\{A,A'\}=0=\{B,B'\}$).
}From now on, we denote such sets of single-qubit orthogonal spin
observables by
 $\{ X^{(1)},Y^{(1)},Z^{(1)}\}$,
 with the choice  of $X^{(1)},~Y^{(1)},~Z^{(1)}$  orthogonal but further
 arbitrary,}
\forget{ It is well known that for all such observables and all
separable states, the Bell
 inequality in the form derived by Clauser, Horne, Shimony and Holt (CHSH) \cite{chsh}
holds:
 \begin{equation}
  |\langle XX + XY + YX -
YY\rangle|  \leq 2,~~\forall\rho_{} \in {\cal
D}^{2\textrm{-sep}}_2~.
 \label{1}
 \end{equation}
}

So, let  two triples of locally orthogonal observables $\{
X^{(1)}_a,Y^{(1)}_a,Z^{(1)}_a\}$ and $\{
X^{(1)}_b,Y^{(1)}_b,Z^{(1)}_b\}$, be given, where $a,b$ label the different
qubits. We introduce two sets of four two-qubit operators on
$\mathcal{H}=\mathbb{C}^2\otimes\mathbb{C}^2$, labeled by the
subscript  $x= 0,1$:
\begin{align}
X_0^{(2)}&:=   \half (X^{(1)}X^{(1)} - Y^{(1)}Y^{(1)})
 & X_1^{(2)}&:=   \half (X^{(1)}X^{(1)} + Y^{(1)}Y^{(1)})\nn\\
Y_0^{(2)}&:=   \half (Y^{(1)}X^{(1)} + X^{(1)}Y^{(1)} )
&{Y}_1^{(2)}&:=   \half (Y^{(1)}X^{(1)} - X^{(1)}Y^{(1)} ) \nn\\
 Z_0^{(2)} &:=  \half (Z^{(1)}I^{(1)} + I^{(1)}Z^{(1)})
 &{Z}_1^{(2)}& :=  \half ( Z^{(1)}I^{(1)} - I^{(1)}Z^{(1)} ) \nn\\
I_0^{(2)} &:= \half (I^{(1)}I^{(1)} + Z^{(1)}Z^{(1)})
 &{I}_1^{(2)}& := \half (I^{(1)}I^{(1)} - Z^{(1)}Z^{(1)}).
 \label{set2}
   \end{align}
 Here,  the superscript label
indicates that we are dealing with two-qubit operators. Later on,
$X_x^{(2)}$
 will sometimes be notated as $X_{x,ab}^{(2)}$,  and similarly for $Y_x^{(2)}$,
 $Z_x^{(2)}$ and $I_x^{(2)}$. This more extensive labeling will prove
 convenient for the multiqubit generalization.
  Note that $(X^{(2)}_x)^2 = (Y^{(2)}_x)^2 = (Z^{(2)}_x)^2 = (I^{(2)}_x)^2 =I^{(2)}_x$
 for $x=0,1$,  and that all eight operators mutually anti-commute.
Furthermore, if the orientations of the two triples are the same,
these two sets  form
 representations of the generalized
Pauli group, i.e., they have the same commutation relations as the
Pauli matrices on $\C^2$, i.e.:
$[X_x^{(2)},Y_x^{(2)}]=2iZ_x^{(2)}$, etc.\, and \beq
\av{X_x^{(2)}}^2+\av{Y_x^{(2)}}^2+\av{Z_x^{(2)}}^2\leq\av{I_x^{(2)}}^2,~~~
x\in \{0,1\}, \label{altijd} \enq with equality only for pure
states.  \forget{ Note that we can rewrite the CHSH inequality
(\ref{1}) in terms of these observables as: $|\av{X_0^{(2)} +
Y_0^{(2)}}|\leq 1, ~ \forall\rho_{} \in {\cal
D}^{2\textrm{-sep}}_2$.}

Assume for the moment that the two-qubit state is pure and
separable. We may thus write $\rho = \ket{\Psi}\bra{\Psi}$, where
$\ket{\Psi} = \ket{\psi} \ket{\phi}$, to obtain:
\begin{align}
\av{X_0^{(2)}}_\Psi^2 + \av{Y_0^{(2)}}_\Psi^2 =
\av{X_1^{(2)}}_\Psi^2 + \av{Y_1^{(2)}}_\Psi^2  &=
 \frac{1}{4}\big( \av{X_a^{(1)}}_\psi^2 + \av{Y_a^{(1)}}_\psi^2\big)
  \big(\av{X_b^{(1)}}_\phi^2 + \av{Y_b^{(1)}}_\phi^2\big) \nn\\
&=\frac{1}{4} \big(\av{I_a^{(1)}} -
\av{Z_a^{(1)}}_\psi^2\big)\big(\av{I_b^{(1)}} -
\av{Z_b^{(1)}}_\phi^2\big) \nn\\  &= \av{I_0^{(2)}}_{{\Psi}}^2 -
\av{Z_0^{(2)}}_{{\Psi}}^2 = \av{I_1^{(2)}}_{{\Psi}}^2 -
\av{Z_1^{(2)}}_{{\Psi}}^2.\label{newa} 
\end{align} 
This result for pure
separable states can be extended to any  mixed separable state
$\rho_{} \in {\cal D}^{2\textrm{-sep}}_2$ by noting that the
density operator of any such state is a convex combination of the
density operators for pure product-states, i.e. $\rho = \sum_j p_j
\ket{\Psi_j}\bra{\Psi_j}$, with $\ket{\Psi_j} =\ket{\psi_j}
\ket{\phi_j}$,  $p_j\geq 0$ and $\sum_j p_j=1$.  We may thus write
for such states:
\begin{align}    \label{convexA} \sqrt{\av{X_x^{(2)}}^2 + \av{Y_x^{(2)}}^2} \leq \sum_{j} p_j
\sqrt{\av{X_x^{(2)} }^2_j +  \av{X_x^{(2)}}^2_j }& = \sum_{j} p_j
\sqrt{\av{I_y^{(2)} }^2_j - \av{Z_y^{(2)}}^2_j } \nn\\&\leq
\sqrt{\av{I_y^{(2)}}^2 - \av{Z_y^{(2)}}^2},
 ~~\forall\rho_{} \in {\cal
D}^{2\textrm{-sep}}_2, ~x,y =0,1.
\end{align}
Here, $\av{\cdot}_j$ denotes  an expectation value in the state
$\ket{\Psi_j}$. \forget{(e.g., $\av{Z_a^{(1)}}_j:=\bra{\Psi_j}
Z_a^{(1)}\otimes\1\ket{\Psi_j}$) and
$\av{Z_a^{(1)}}:=\av{Z_a^{(1)}\otimes\1}$. } The first inequality
follows because
  $\sqrt{\av{X_x^{(2)}}^2 + \av{Y_x^{(2)}}^2  }$ are
convex functions of $\rho$  for all $x$ and the second because
 $\sqrt{\av{I_y^{(2)}}^2 - \av{Z_y^{(2)}}^2}$
 are  concave in $\rho$ for all $y$.
\forget{
 Thus, we obtain for all separable states and
locally orthogonal triples:
\begin{align} \av{X_x^{(2)}}^2 + \av{Y_x^{(2)}}^2   \leq
\frac{1}{4}(\av{I_a^{(1)}} - \av{Z^{(1)}_a}^2)(\av{I_b^{(1)}} - \av{Z^{(1)}_b}^2)
,~~~\forall\rho_{} \in {\cal
D}^{2\textrm{-sep}}_2 .\label{new}
\end{align}}
As shown in \cite{uffseev} the right-hand side of this inequality is bounded by $1/2$, which follows by considering the equalities of \eqref{newa}.
However, for  entangled states (e.g., for the Bell states
$\ket{\phi^{\pm}}=(\ket{00} \pm \ket{11})/\sqrt{2}$ and
$\ket{\psi^{\pm}}=(\ket{01} \pm \ket{10})/\sqrt{2}$) the left-hand
side can attain the value of 1. Hence, inequality (\ref{convexA})
provides a nontrivial bound for separable states, and thus a
criterion for testing entanglement.

\forget{ Note that using only two correlation terms (instead of
four as in (\ref{1})) one can already find a separability
criterion for the case of two qubits (also noted in Ref.\
\cite{tothguhne1}) implies: \beq \forall\rho_{} \in {\cal
D}^{2\textrm{-sep}}_2\,:\, |\av{X_x^{(2)}}|\leq 1/2 . \enq
\forget{ Violation of this inequality thus gives an entanglement
criterion, i.e., if $|\av{X_x^{(2)}}|> 1/2$ then $\rho$ is
entangled. In fact,} A maximally entangled state can give rise to
$|\av{X_x^{(2)}}|=1$ a factor two higher than for separable
states. The same of course holds for the choice $Y_x^{(2)}$. }

\forget{ It is also instructive to relate the results obtained so
far to the original Bell-CHSH inequality (\ref{1}). Since
$(\av{X_0^{(2)} + Y_0^{(2)}})^2 +(\av{X_0^{(2)} - Y_0^{(2)}})^2=2
(\av{X_0^{(2)}}^2 + \av{Y_0^{(2)}}^2)$ we obtain that
$|\av{X_0^{(2)} + Y_0^{(2)}}|\leq \sqrt{1/2}$, which strengthens
the original CHSH inequality by a factor $\sqrt{2}$.
 Usually Bell inequalities are considered
interesting only if there exist a certain quantum state that
violates it for certain measurement settings. However, we see here
that it is also very interesting to ask not what quantum states
violate a certain Bell inequality, but what quantum states cannot
violate such a Bell inequality, and by what factor.  This is
especially so if this factor is lower than the bound for LHV
models since  then LHV correlations would exist which cannot be
reproduced by separable quantum states,  cf.\ Ref.\
\cite{uffseev}. In fact, section \ref{Nqubitsection} shows that
this divergence between  the correlations obtainable by LHV models
and by separable quantum states generalizes to the $N$-qubit case
and increases exponentially with $N$.}

\forget{  Let us first temporarily assume the state to be pure and
separable, $\ket{\Psi} =\ket{\psi}\ket{\phi}$.  Then we obtain
from (\ref{newa})
\begin{align}
\av{X_0^{(2)}}_{{\Psi}}^2 + \av{Y_0^{(2)}}_{\Psi}^2 =\av{X_1^{(2)}}_{{\Psi}}^2
+ \av{Y_1^{(2)}}_{{\Psi}}^2  \label{XY},
\end{align}
and similarly: \begin{align} \av{I_0^{(2)}}_{{\Psi}}^2 -
\av{Z_0^{(2)}}_{{\Psi}}^2 = \frac{1}{4} \left( \av{I_a^{(1)}}_\psi
-\av{Z_a^{(1)}}_\psi^2 \right) \left( \av{I_b^{(1)}}_\phi -
\av{Z_b^{(1)}}_\phi^2 \right) = \av{I_1^{(2)}}_{{\Psi}}^2 -
\av{Z_1^{(2)}}_{{\Psi}}^2  \label{IZ}.\end{align}} \forget{
 In view of
(\ref{newa}) we conclude that for all pure separable states all
expressions in the equations (\ref{XY}) and (\ref{IZ}) are equal
to each other.   Of course, this conclusion does not hold for
mixed separable states. However,  $\sqrt{\av{X_0^{(2)}}^2 +
\av{Y_0^{(2)}}^2}$ and   $\sqrt{\av{X_1^{(2)}}^2 +
\av{Y_1^{(2)}}^2}$ are convex functions of $\rho$ whereas
 the three expressions
 $\sqrt{\av{I_0^{(2)}}^2 - \av{Z_0^{(2)}}^2}$,
 $\sqrt{\frac{1}{4} ( \av{I_a^{(1)}} -\av{Z_a^{(1)}}^2 )
(\av{I_b^{(1)}} - \av{I_a^{(1)}}^2 )}$
 and $\sqrt{\av{I_1^{(2)}}^2 - \av{Z_1^{(2)}}^2}$
 are all concave in $\rho$.
 Therefore we can repeat
 a similar
 chain of reasoning as in (\ref{convexA})  to  obtain  the following inequalities, which are valid
for all mixed separable states:
\begin{align} \max\left \{
\begin{array}{c}
\av{X_0^{(2)}}^2 + \av{Y_0^{(2)}}^2 \\
\av{X_1^{(2)}}^2
+ \av{Y_1^{(2)}}^2  \end{array} \right\}
\leq \min
\left\{ \begin{array}{c}
\av{I_0^{(2)}}^2 - \av{Z_0^{(2)}}^2 \\
\frac{1}{4} \left( \av{I_a^{(1)}} -\av{Z_a^{(1)}}^2 \right)
\left(\av{I_b^{(1)}} - \av{Z_b^{(1)}}^2 \right)\\
\av{I_1^{(2)}}^2 - \av{Z_1^{(2)}}^2
\end{array}
\right\} \leq\frac{1}{4}, ~~~\forall\rho_{} \in {\cal
D}^{2\textrm{-sep}}_2.
\label{mixsep}\end{align}

 }
 In other words, for all separable 2-qubit
states one has:
\forget{\begin{align} %\left.
\max \left\{ \begin{array}{c}\av{X_0^{(2)}}^2 + \av{Y_0^{(2)}}^2\\
\av{X_1^{(2)}}^2 + \av{Y_1^{(2)}}^2\end{array} \right\}
\leq
\min \left\{ \begin{array}{c}
\av{I_0^{(2)}}^2 - \av{Z_0^{(2)}}^2\\
\av{I_1^{(2)}}^2 - \av{Z_1^{(2)}}^2  \end{array}
 \right\}\leq \frac{1}{4}, ~~\forall\rho_{} \in {\cal
D}^{2\textrm{-sep}}_2. \label{mixsep2}
\end{align}}
\begin{align} %\left.
\max_{x\in \{0,1\}}\av{X_x^{(2)}}^2 + \av{Y_x^{(2)}}^2
\leq
\min_{x\in\{0,1\}} \av{I_x^{(2)}}^2 - \av{Z_x^{(2)}}^2
\leq \frac{1}{4}, ~~\forall\rho_{} \in {\cal
D}^{2\textrm{-sep}}_2. \label{mixsep2}
\end{align}
In fact, the validity of the inequalities (\ref{mixsep2}) for all
orthogonal triples $\{ X^{(1)}_a,Y^{(1)}_a,Z^{(1)}_a\}$ and $\{
X^{(1)}_b,Y^{(1)}_b,Z^{(1)}_b\}$ provides a necessary and
sufficient condition for separability for two-qubit states, pure
or mixed. (See \cite{uffseev} for a proof).

Note that, depending on whether the orientation of the triples of
local orthogonal observables is the same or not, the inequalities
on the left-hand side of (\ref{mixsep2}) (leaving out the upperbound $1/4$) may be simplified. If we
choose the orientations for both parties to be the same, then the
interesting separability inequalities in (\ref{mixsep2}) are
$\av{X_0^{(2)}}^2 + \av{Y_0^{(2)}}\leq \av{I_1^{(2)}}^2 -
\av{Z_1^{(2)}}^2$ and $\av{X_1^{(2)}}^2 + \av{Y_1^{(2)}} \leq
\av{I_0^{(2)}}^2 - \av{Z_0^{(2)}}^2$, whereas the other
inequalities in (\ref{mixsep2}) become trivially true (cf.\
(\ref{altijd})).
 Choosing the orientations to be different reverses this verdict.

To conclude this section we give an explicit form of the
separability inequalities (\ref{mixsep2}) by choosing the
Pauli matrices $\{\sigma_x,\sigma_y,\sigma_z\}$ for both triples
$\{ X^{(1)}_a,Y^{(1)}_a,Z^{(1)}_a\}$ and $\{
X^{(1)}_b,Y^{(1)}_b,Z^{(1)}_b\}$. This choice enables us to write
the inequalities (\ref{mixsep2}) in terms of the density matrix
elements on the standard  $z$-basis
$\{\ket{00},\ket{01},\ket{10},\ket{11}\}$, labeled here as  $\{\ket{1},\ket{2},\ket{3},\ket{4}\}$.
  This choice of observables yields  $\av{
X_0^{(2)}}=2\mathrm{Re}\, \rho_{1,4}$,  $\av{
Y_0^{(2)}}=-2\mathrm{Im}\, \rho_{1,4}$, $\av{ I_0^{(2)}}=\rho_{1,1}+
\rho_{4,4}$,  $\av{ Z_0^{(2)}}=\rho_{1,1} -\rho_{4,4}$,
 $\av{X_1^{(2)}}=2\mathrm{Re}\, \rho_{2,3}$,
 \forget{ $Y_1^{(2)}=
-i\ket{01}\bra{10} +i\ket{10}\bra{01}$,}
 $\av{ Y_1^{(2)}}=-2\mathrm{Im}\, \rho_{2,3}$, \forget{, $I_1^{(2)}=
\ket{01}\bra{01} +\ket{10}\bra{10}$,} $\av{ I_1^{(2)}}=\rho_{2,2}+
\rho_{3,3}$, \forget{ $Z_1^{(2)}= \ket{01}\bra{01}
-\ket{10}\bra{10}$,} $\av{ Z_1^{(2)}}=\rho_{2,2} -\rho_{3,3}$.
So, in this choice, we can write  (\ref{mixsep2}) as:
\begin{align}
\max\{ |\rho_{1,4}|^2,\, |\rho_{2,3}|^2 \} \leq \min\{
\rho_{1,1}\rho_{4,4},\, \rho_{2,2} \rho_{3,3} \}\leq \frac{1}{16}, ~~~~   \rho_{} \in {\cal
D}^{2\textrm{-sep}}_2.
 \label{2sepmatrix} 
\end{align}

In the form (\ref{2sepmatrix}), it is easy to compare the result
to the separability conditions reviewed in subsection  II.B.  Assume for simplicity that $|\rho_{1,4}|$ is the largest of all
the antidiagonal elements  $|\rho_{j\bar{\jmath}}|$. Then, for
$\rho_{} \in {\cal D}^{2\textrm{-sep}}_2$, and using $\av{M^{(2)}}^2 +\av{M'^{(2)}}^2=8(\av{X_0^{(2)}}^2+\av{Y_0^{(2)}}^2)$ the Mermin-type separability inequality \eqref{quadraticN} becomes $|\rho_{1,4}|^2\leq 1/16$, which is equivalent to the Laskowski-\.Zukowski condition $|\rho_{1,4}|\leq 1/4$; the
fidelity/D\"ur-Cirac conditions read: $2|\rho_{1,4}|\leq
\rho_{2,2} +\rho_{3,3}$; and the condition (\ref{2sepmatrix}):
 $|\rho_{1,4}|^2\leq \rho_{2,2}\rho_{3,3}$. Using the trivial inequality $(\sqrt{\rho_{22}} -\sqrt{\rho_{33}})^2\geq0\Longleftrightarrow  2\sqrt{ \rho_{2,2}\rho_{3,3}}\leq \rho_{2,2} +\rho_{3,3}$, we can then write the following chain of inequalities:
\beq
  4|\rho_{1,4}|- (\rho_{1,1}+\rho_{4,4}) \overset{A}{\leq}2 |\rho_{1,4}| \overset{\textrm{sep}}{\leq} 2\sqrt{\rho_{2,2}\rho_{3,3}}  \overset{A}{\leq}\rho_{2,2} + \rho_{3,3}\,,
  \label{inequ2}
  \eeq
where we used the symbols  $\overset{A}{\leq}$
and $\overset{\textrm{sep}}{\leq}$ to denote inequalities that hold for all states, and for the separability condition (\ref{2sepmatrix}) respectively.

The Laskowski-\.Zukowski condition is then recovered by comparing the first and fourth expressions in this chain, the fidelity/ D\"ur-Cirac conditions by comparing the second and fourth expression, and a new condition -- not previously mentioned -- can be obtained by
comparing the first and third term, whereas condition (\ref{2sepmatrix}), i.e.
the comparison between the second and third expression in (\ref{inequ2}), is the strongest inequality in this chain, and thus implies and strengthens all of these other conditions.

\subsection{Three-qubit case}

 We now derive separability conditions that distinguish
the $10$ classes in the $3$-qubit classification of section
\ref{generalksep} by generalizing the method of section
\ref{twoqubitsection}. To begin with, define four sets of
three-qubit observables from the two-qubit operators  (\ref{set2})
.
\begin{align}
X_0^{(3)} &:=\frac{1}{2}\,(X^{(1)}   X^{(2)}_0 -Y^{(1)}  Y^{(2)}_0)
&{X}_1^{(3)}
&:=\frac{1}{2}\,(X^{(1)}   {X}^{(2)}_0 +Y^{(1)}  Y^{(2)}_0)\nn
\\
Y_0^{(3)}  &:= \frac{1}{2}\,(Y^{(1)}  X^{(2)}_0+X^{(1)}   Y^{(2)}_0) &
{Y}_1^{(3)}
&:=\frac{1}{2}\,(Y^{(1)}   X^{(2)}_0-X^{(1)}  Y^{(2)}_0 )\nn
\\
Z_0^{(3)}  &:= \frac{1}{2}\,(Z^{(1)}   I^{(2)}_0+I^{(1)}  Z^{(2)}_0) &{Z}_1^{(3)}
&:=\frac{1}{2}\,(Z^{(1)}  I^{(2)}_0-I^{(1)}   Z^{(2)}_0 )\nn
\\
I_0^{(3)}&:=\frac{1}{2}\,(I^{(1)}   I^{(2)}_0 +Z^{(1)}  Z^{(2)}_0) &{I}_1^{(3)}
&:=\frac{1}{2}\,(I^{(1)}   {I}^{(2)}_0 -Z^{(1)}  Z^{(2)}_0)\nn
\\\nn\\
X_2^{(3)}  &:=\frac{1}{2}\,(X^{(1)}   X^{(2)}_1 -Y^{(1)}
Y^{(2)}_1) &
X_3^{(3)} &:=\frac{1}{2}\,(X^{(1)}
 X^{(2)}_1 +Y^{(1)} Y^{(2)}_1)\nn
\\
Y^{(3)}_2 &:=\frac{1}{2}\,(Y^{(1)}   {X}^{(2)}_1 +X^{(1)}
{Y}^{(2)}_1 ) &Y_3^{(3)} &:=\frac{1}{2}\,(Y^{(1)}   {X}^{(2)}_1-X^{(1)}
{Y}^{(2)}_1 )\nn
\\
Z_2^{(3)} &:=\frac{1}{2}\,(Z^{(1)}  {I}^{(2)}_1 +I^{(1)}
{Z}^{(2)}_1) &Z_3^{(3)} &:=\frac{1}{2}\,(Z^{(1)}
  {I}^{(2)}_1-I^{(1)}  {Z}^{(2)}_1 )\nn
\\
I_2^{(3)} &:=\frac{1}{2}\,(I^{(1)}   {I}^{(2)}_1 +Z^{(1)}
 Z^{(2)}_1)
 &I^{(3)}_3 &:=\frac{1}{2}\,(I^{(1)}   {I}^{(2)}_1
 -Z^{(1)}  {Z}^{(2)}_1)
,\label{N3operators}
\end{align}
where $X^{(1)}X^{(2)}_{0}= X^{(1)}_a \otimes X^{(2)}_{0,bc}$,
etc.,  $a,b,c$ label the three qubits. In analogy to
 the two-qubit case, we note that all these operators anticommute and that if the orientations of the triples  for each qubit
are the same, the operators in (\ref{N3operators}) yield
representations of the generalized Pauli group:
$[X_x^{(3)},Y_x^{(3)}]=2iZ_x^{(3)}$, for    $x=0,1,2,3$. For
convenience, we will indeed assume these orientations to be the
same, unless noted otherwise. Choosing orientations differently
would yield similar separability
 conditions, in the same vein as in the previous section.
Under this choice we have, for all $k$,   \beq
\av{X_x^{(3)}}^2+\av{Y_x^{(3)}}^2+\av{Z_x^{(3)}}^2\leq\av{I_x^{(3)}}^2,~~~\forall
\rho \in {\cal D}_N^{k\textrm{-sep}} \label{eenheid} \eeq
 with equality only for pure states.

We now derive conditions for the different levels and classes of
the partial separability classification. Most of the proofs are by
straightforward generalization of the method of the previous
section and these will be omitted.

Suppose first that the three-qubit state  is pure and separable
under split $a$-$(bc)$. From the definitions  (\ref{N3operators})
we obtain: 
\begin{eqnarray}
 \av{X_0^{(3)} }^2 +\av{Y_0^{(3)} }^2=&
\frac{1}{4}\,(\,\av{X^{(1)}_a}^2+\av{Y^{(1)}_a}^2\,)\,
(\,\av{X^{(2)}_{0,bc}}^2 +\av{Y^{(2)}_{0,bc}}^2\,)&=
 \av{X_1^{(3)} }^2 +\av{Y_1^{(3)} }^2 =
\nn\\
\av{I_0^{(3)} }^2 -\av{Z_0^{(3)}
}^2=&\frac{1}{4}\,(\,\av{I^{(1)}_a}^2
-\av{Z^{(1)}_a}^2\,)\,(\,\av{I^{(2)}_{0,bc}}^2
-\av{Z^{(2)}_{0,bc}}^2\,)&=
 \av{I_1^{(3)} }^2 -\av{Z_1^{(3)} }^2,
\label{2_sep_a_bcX} 
\\
\av{X_2^{(3)} }^2 +\av{Y_2^{(3)}
}^2=&\frac{1}{4}\,(\,\av{X^{(1)}_a}^2
+\av{Y^{(1)}_a}^2\,)\,(\,\av{X^{(2)}_{1,bc}}^2
+\av{Y^{(2)}_{1,bc}}^2\,) &= \av{X_3^{(3)} }^2 +\av{Y_3^{(3)} }^2
=
\nn\\
\av{I_2^{(3)} }^2 -\av{Z_2^{(3)} }^2=&\frac{1}{4}\,(\,\av{I^{(1)}_a}^2
-\av{Z^{(1)}_a}^2\,)\,(\,\av{I^{(2)}_{1,bc}}^2 -\av{Z^{(2)}_{1,bc}}^2\,)&=
 \av{I_3^{(3)} }^2 -\av{Z_3^{(3)} }^2.\label{2_sep_a_bcI}
 \end{eqnarray}
  Similarly, for pure states  that are separable under
split $b$-$(ac)$, we obtain  analogous equalities by interchanging
the labels  $x=1$ and $x=3$ (denoted as $1 \leftrightarrow
3$); and for split $c$-$(ab)$ by $1\leftrightarrow 2$.

Of course,  these equalities  hold for pure states only, but by
 the convex analysis of section \ref{twoqubitsection} we obtain from (\ref{2_sep_a_bcX}, \ref{2_sep_a_bcI}) inequalities for all mixed states that are biseparable under
the split $a$-$(bc)$:
\begin{align}\begin{array}{clcl}
\max\limits_{x \in \{0,1\}} \av{X_x^{(3)} }^2 +\av{Y_x^{(3)} }^2 \leq
\min\limits_{x \in \{ 0,1\}} \av{I_x^{(3)} }^2 -\av{Z_x^{(3)} }^2
\leq \frac{1}{4}
\\
\max\limits_{x \in \{2,3\}} \av{X_x^{(3)} }^2 +\av{Y_x^{(3)} }^2 \leq
\min\limits_{x \in \{ 2,3\}} \av{I_x^{(3)} }^2 -\av{Z_x^{(3)} }^2 \leq
\frac{1}{4}
\end{array}, ~~ ~\forall\rho \in {\cal D}_3^{a\textrm{-}(bc)}.
  \label{ineq_2sep}
  \end{align}
For states that are biseparable under split $b$-$(ac)$ the
analogous inequalities with $1\leftrightarrow 3$ hold, i.e.,
\begin{align}
\begin{array}{clcl}\max\limits_{x \in \{0,3\}} \av{X_x^{(3)} }^2 +\av{Y_x^{(3)} }^2 \leq
\min\limits_{x \in \{ 0,3\}} \av{I_x^{(3)} }^2 -\av{Z_x^{(3)} }^2
\leq \frac{1}{4}\\
\max\limits_{x \in \{1,2\}} \av{X_x^{(3)} }^2 +\av{Y_x^{(3)} }^2 \leq
\min\limits_{x \in \{ 1,2\}} \av{I_x^{(3)} }^2 -\av{Z_x^{(3)} }^2 \leq
\frac{1}{4}
\end{array},  ~~ ~\forall\rho \in {\cal D}_3^{b\textrm{-}(ac)}.
  \label{ineq_2sepb}
  \end{align}
 and for
the split $c$-$(ab)$ we need to replace $1 \leftrightarrow 2$:
\begin{align}
\begin{array}{clcl}\max\limits_{x \in \{0,2\}} \av{X_x^{(3)} }^2 +\av{Y_x^{(3)} }^2 \leq
\min\limits_{x \in \{ 0,2\}} \av{I_x^{(3)} }^2 -\av{Z_x^{(3)} }^2
\leq \frac{1}{4}
\\
\max\limits_{x \in \{1,3\}} \av{X_x^{(3)} }^2 +\av{Y_x^{(3)} }^2 \leq
\min\limits_{x \in \{ 1,3\}} \av{I_x^{(3)} }^2 -\av{Z_x^{(3)} }^2 \leq
\frac{1}{4}
\end{array},
 ~~ ~\forall\rho \in {\cal D}_3^{c\textrm{-}(ab)}.
  \label{ineq_2sepc}
% \label{N3ok}
  \end{align}

A general biseparable state $\rho_{} \in {\cal
D}^{2\textrm{-sep}}_3$  is a  convex mixture of states that are
separable under some bipartite split, i.e.,
 $ \rho=p_1 \rho_{a\textrm{-}(bc)}+p_2 \rho_{b\textrm{-}(ac)}+p_3 \rho_{c\textrm{-}(ab)}$ with
$\sum_{j=1}^3 p_j=1$.
 Since $\sqrt{\av{{X^{(3)}_0}}^2 +\av{{Y^{(3)}_0}}^2}$ is convex in $\rho$ we get from (\ref{ineq_2sep}- \ref{ineq_2sepc})
 for such a state:
\begin{align}\label{convex2sep3}
\sqrt{\av{{X^{(3)}_0}}^2 +\av{{Y^{(3)}_0}}^2} & \leq p_1\sqrt{
\av{{X^{(3)}_0}}^2_{ \rho_{a\textrm{-}(bc)}} +
\av{{Y^{(3)}_0}}^2_{ \rho_{a\textrm{-}(bc)}}}+ p_2\sqrt{
\av{X^{(3)}_0}^2_{ \rho_{b\textrm{-}(ac)}} +\av{Y^{(3)}_0}^2_{ \rho_{b\textrm{-}(ac)}}}+
p_3\sqrt{\av{X^{(3)}_0}^2_{ \rho_{c\textrm{-}(ab)}} +\av{Y^{(3)}_0}^2_{ \rho_{c\textrm{-}(ab)}}}\nn\\
&\leq p_1 \sqrt{\av{I^{(3)}_1}^2_{ \rho_{a\textrm{-}(bc)}}
-\av{Z^{(3)}_1}^2_{ \rho_{a\textrm{-}(bc)}}}+ p_2\sqrt{
\av{I^{(3)}_3}^2_{ \rho_{b\textrm{-}(ac)}} - \av{Z^{(3)}_3}^2_{ \rho_{b\textrm{-}(ac)}}}+
p_3 \sqrt{\av{I^{(3)}_2}^2_{ \rho_{c\textrm{-}(ab)}} -
\av{Z^{(3)}_2}^2_{ \rho_{c\textrm{-}(ab)}}}.
\end{align}
Here $\av{\cdot}_{ \rho_{a\textrm{-}(bc)}}$ means taking the
expectation value in the state $ \rho_{a\textrm{-}(bc)}$, etc.
Analogous bounds hold for the expressions $\sqrt{\av{X_x^{(3)} }^2
+\av{Y_x^{(3)} }^2}$ for  $x=1,2,3$.
%In general the fractions $p_j$ may be unknown.

From the numerical upper bounds in the conditions
(\ref{ineq_2sep}- \ref{ineq_2sepc}) it is easy to obtain a first
biseparability  condition: \beq\label{first3} \av{X_x^{(3)} }^2
+\av{Y_x^{(3)} }^2\leq 1/4, ~~~ \forall\rho_{} \in {\cal
D}^{2\textrm{-sep}}_3, ~~~x\in\{0,1,2,3\}. \eeq This is
equivalent to the Laskowski-\.Zukowski condition
(\ref{antidiagonal}) for $k=2$, as will be shown below.
 However, a stronger condition can be obtained  by noting that $\sqrt{\av{I^{(3)}_y}^2
-\av{Z^{(3)}_y}^2}$ is concave in $\rho$ so that
\begin{align}
p_1\sqrt{
\av{{I^{(3)}_y}}^2_{ \rho_{a\textrm{-}(bc)}} -
\av{{Z^{(3)}_y}}^2_{ \rho_{a\textrm{-}(bc)}}}+ p_2\sqrt{
\av{I^{(3)}_y}^2_{ \rho_{b\textrm{-}(ac)}} -\av{Z^{(3)}_y}^2_{ \rho_{b\textrm{-}(ac)}}}+
p_3\sqrt{\av{I^{(3)}_y}^2_{ \rho_{c\textrm{-}(ab)}} -\av{Z^{(3)}_y}^2_{ \rho_{c\textrm{-}(ab)}}}\leq
\sqrt{\av{I^{(3)}_y}^2 -\av{Z^{(3)}_y}^2}.
\label{conc}
\end{align}
After taking a sum over $y\neq x$ in (\ref{conc}),  the left hand
side of (\ref{conc}) is larger than the right hand side of
(\ref{convex2sep3}). This yields a stronger condition for
biseparability  of  $3$-qubit states \beq\label{2sep3}
\sqrt{\av{X^{(3)}_{x}}^2 +\av{Y^{(3)}_{x}}^2} \leq\sum_{y\neq x}
\sqrt{ \av{I^{(3)}_{y}}^2 -\av{Z^{(3)}_{y}}^2}, ~~~ \forall\rho_{}
\in {\cal D}^{2\textrm{-sep}}_3, ~~~x,y\in\{0,1,2,3\}. \eeq That
(\ref{2sep3}) is indeed a stronger than (\ref{first3}) will be
shown below using the density matrix representation of this
condition.
 If one
would alter the orientation of the orthogonal triple of
observables for a certain qubit, then the right-hand side of
(\ref{2sep3}) changes by adding either $1$, $2$ or $3$ (modulo
$3$) to $x$  in the sum on the right hand side, depending on for
which qubit the orientation was changed.

Next, consider the case  of a  $3$-separable state, $\rho \in
{\cal D}_3^{3\textrm{-sep}}$.  One might then use the fact that
this split is contained in all three bipartite splits $a$-$(bc)$,
$b$-$(ac)$ and $c$-$(ab)$ to conclude that the  inequalities
(\ref{ineq_2sep}, \ref{ineq_2sepb}, \ref{ineq_2sepc}) must hold
simultaneously. Thus,  3-separable states must obey: \beq \max_{x}
\{ \av{X^{(3)}_{x}}^2 +\av{Y^{(3)}_{x}}^2 \} \leq \min_{x} \{
\av{I^{(3)}_{x}}^2 -\av{Z^{(3)}_{x}}^2\} \leq
\frac{1}{4},~~~\forall \rho \in {\cal D}_3^{3\textrm{-sep}}
\label{3sep3}.
 \eeq
However, a more stringent condition
 holds by virtue of
the following equalities  for  pure $3$-separable states: 
\begin{align}
\av{{X^{(3)}_0}}^2 +\av{{Y^{(3)}_0}}^2 &=
\frac{1}{16}\,(\,\av{X^{(1)}_a}^2+\av{Y^{(1)}_a}^2\,)
\,(\,\av{X^{(1)}_b}^2+\av{Y^{(1)}_b}^2\,)\,(\,\av{X^{(1)}_c}^2+\av{Y^{(1)}_c}^2\,)
\nn\\
&=\av{X^{(3)}_1}^2 +\av{Y^{(3)}_1}^2=\av{ X^{(3)}_2}^2 +
\av{Y^{(3)}_2}^2=\av{ X^{(3)}_3}^2 +
\av{Y^{(3)}_3}^2,
\\
\av{{I^{(3)}_0}}^2 -\av{{Z^{(3)}_0}}^2 &=
\frac{1}{16}\,(\,\av{I^{(1)}_a}^2-\av{Z^{(1)}_a}^2\,)
\,(\,\av{I^{(1)}_b}^2-\av{Z^{(1)}_b}^2\,)\,(\,\av{I^{(1)}_c}^2-\av{Z^{(1)}_c}^2\,)
\nn\\
&=\av{I^{(3)}_1}^2 -\av{Z^{(3)}_1}^2=\av{ I^{(3)}_2}^2 -
\av{Z^{(3)}_2}^2=\av{ I^{(3)}_3}^2 -
\av{Z^{(3)}_3}^2.\label{completesep} 
\end{align} From these equalities
for pure states it is easy to obtain, by a convexity argument
similar to previous cases, an upper bound of $1/16$
instead of $1/4$ in (\ref{3sep3}):
  \beq \max_{x}
\{ \av{X^{(3)}_{x}}^2 +\av{Y^{(3)}_{x}}^2 \} \leq \min_{x} \{
\av{I^{(3)}_{x}}^2 -\av{Z^{(3)}_{x}}^2\} \leq
\frac{1}{16},~~~\forall \rho \in {\cal D}_3^{3\textrm{-sep}}
\label{3sep3b}.
 \eeq\forget{Note that by the same convex
analysis of section \ref{twoqubitsection} these equalities allow
for direct derivation of the inequality (\ref{3sep3}) without
using the biseparability inequalities (\ref{ineq_2sep}).}

 We have thus  obtained different conditions for each of the 10 classes
in the full separability classification of three qubits,
summarized in table \ref{table1}.
\begin{table}[!h]
\begin{tabular}{|l|l|} \hline
Class &  Separability conditions \\ \hline
1 &  (\ref{eenheid}) \\
2.1& (\ref{2sep3})\\
 2.2 & (\ref{ineq_2sep})  \\
 2.3 & (\ref{ineq_2sepb})  \\
 2.4 & (\ref{ineq_2sepc}) \\
 2.5  &  (\ref{ineq_2sep}) \&  (\ref{ineq_2sepb}) but not (\ref{ineq_2sepc})\\
2.6  &   (\ref{ineq_2sep}) \&  (\ref{ineq_2sepc}) but not
(\ref{ineq_2sepb})
 \\
2.7 &  (\ref{ineq_2sepb}) \& (\ref{ineq_2sepc}) but not (\ref{ineq_2sep}) \\
2.8 &  ((\ref{ineq_2sep}) \& (\ref{ineq_2sepb})\& (\ref{ineq_2sepc}))$ \equivto$ (\ref{3sep3}) \\
 3 &   (\ref{3sep3b})\\ \hline
\end{tabular}
\caption{Separability conditions for the 10 classes in the
separability classification of three-qubit states.} \label{table1}
\end{table}
Violations of these partial separability conditions
 %of table \ref{table1}
  give sufficient conditions for particular types of entanglement. For example,  if inequality (\ref{3sep3b}) is violated, then the state  must be in one of the
biseparable classes $2.1$ to $2.8$ or in class $1$, which implies
that the state is at least $2$-partite entangled; \forget{If
inequality (\ref{ineq_2sep}) is violated, then the state cannot be
in classes $2.2$, $2.5$, $2.6$, $2.8$
 and $3$.  If inequality (\ref{ineq_2sepb}) is violated,  the state cannot be in
classes $2.3$, $2.5$, $2.7$, $2.8$ and $3$. If inequality
(\ref{ineq_2sepc}) is violated, then the state cannot be in
classes $2.4$, $2.6$, $2.7$, $2.8$ and $3$. If either two of the
three inequalities obtained from (\ref{ineq_2sep}) is violated
then the state cannot be in classes $2.1$ and $2.2$, or $2.2$ and
$2.4$ or $2.3$ and $2.4$ and neither can it be in classes $2.5$ to
$2.8$ and $3$. If all these inequalities are violated the state
must be in either in class $1$ or class $2.1$.} if (\ref{2sep3})
violated it is in class $1$  and thus fully inseparable (fully
entangled), and so on.

In order to gain further familiarity with the above separability
inequalities, we choose the ordinary Pauli matrices $\{\sigma_x,
\sigma_y, \sigma_z\}$ for the locally orthogonal observables $ \{
X^{(1)}, Y^{(1)}, Z^{(1)}\}$, and formulate them in terms of
density matrix elements in the standard $z$-basis.
Inequalities (\ref{ineq_2sep},\ref{ineq_2sepb},\ref{ineq_2sepc}) now read successively:\\
\begin{align}
\begin{array}{l}
\max\{
 |\rho_{1,8}|^2,
|\rho_{4,5}|^2
 \}
\leq
\min \{\rho_{4,4}\rho_{5,5}
,
\rho_{1,1}\rho_{8,8}
  \}\leq 1/16\\
 \max\{ |\rho_{2,7}|^2
,
|\rho_{3,6}|^2
 \}
\leq
\min\{
 \rho_{2,2}\rho_{7,7}
, \rho_{3,3}\rho_{6,6} \} \leq 1/16
\end{array},~~~\forall\rho \in {\cal
D}_3^{a\textrm{-}(bc)}, \label{a-bc3sep}
 \end{align}\forget{
while biseparability under split $b$-$(ac)$ (i.e., $\rho \in
{\cal D}_3^{b\textrm{-}(ac)}$) gives:}
    \begin{align}
    \begin{array}{l}
    \max\{
 |\rho_{1,8}|^2,
|\rho_{3,6}|^2
 \}
\leq
\min \{\rho_{3,3}\rho_{6,6}
,
\rho_{1,1}\rho_{8,8}
  \}\leq 1/16\\
 \max\{ |\rho_{2,7}|^2
,
|\rho_{4,5}|^2
 \}
\leq
\min\{
 \rho_{2,2}\rho_{7,7}
, \rho_{4,4}\rho_{5,5} \}\leq 1/16\end{array}, ~~~   \forall \rho \in {\cal
D}_3^{b\textrm{-}(ac)}, \label{b-ac3sep}
 \end{align} \forget{
     and under split $c$-$(ab)$ (i.e., $\rho \in {\cal
     D}_3^{c\textrm{-}(ab)}$):}
      \begin{align}\begin{array}{l}
\max\{
 |\rho_{1,8}|^2,
|\rho_{2,7}|^2
 \}
\leq
\min \{\rho_{2,2}\rho_{7,7}
,
\rho_{1,1}\rho_{8,8}
  \}\leq 1/16 \\
 \max\{ |\rho_{3,6}|^2
,
|\rho_{4,5}|^2
 \}
\leq
\min\{
 \rho_{3,3}\rho_{6,6}
, \rho_{4,4}\rho_{5,5} \}\leq 1/16\end{array}, ~~~ \forall \rho \in {\cal
     D}_3^{c\textrm{-}(ab)}. \label{c-ab3sep}
 \end{align}
       For a general biseparable state  we can rewrite
       (\ref{first3}) as:
 \beq
 \max\{|\rho_{1,8}, |\rho_{2,7}|, |\rho_{3,6}|,|\rho_{4,5}| \} \leq 1/4
 ~~~ \forall \rho \in {\cal D}_3^{2\textrm{-sep}}.\eeq It can easily
 be seen that
 this is equivalent to Laskowski-\.Zukowski's condition
(\ref{antidiagonal}) for $k=2$. The condition (\ref{2sep3}) for
biseparability yields: 
\begin{align}
\begin{array}{l}
|\rho_{1,8}| \leq
\sqrt{\rho_{2,2}\rho_{7,7}}+\sqrt{\rho_{3,3}\rho_{6,6}}+\sqrt{\rho_{4,4}\rho_{5,5}}\\
 |\rho_{2,7}| \leq \sqrt{\rho_{1,1}\rho_{8,8}}+
\sqrt{\rho_{3,3}\rho_{6,6}}+\sqrt{\rho_{4,4}\rho_{5,5}}\\
  |\rho_{3,6}|  \leq \sqrt{\rho_{1,1}\rho_{8,8}}+\sqrt{\rho_{2,2}\rho_{7,7}}+\sqrt{\rho_{4,4}\rho_{5,5}}\\
|\rho_{4,5}|\leq\sqrt{ \rho_{1,1}\rho_{8,8}}+
\sqrt{\rho_{2,2}\rho_{7,7}}+\sqrt{\rho_{3,3}\rho_{6,6}}\end{array},~~~\forall\rho \in {\cal
D}_3^{2\textrm{-sep}}\label{matrixbisep3}.\end{align}
Finally,  condition
(\ref{3sep3}) for general $3$-separable states  becomes: \beq
\max\{ |\rho_{1,8}|^2, |\rho_{2,7}|^2, |\rho_{3,6}|^2,
|\rho_{4,5}|^2\}\leq \min  \{ \rho_{1,1}\rho_{8,8},
\rho_{2,2}\rho_{7,7}, \rho_{3,3}\rho_{6,6},
\rho_{4,4}\rho_{5,5}\}\leq \frac{1}{64}, ~~~\forall\rho \in {\cal
D}_3^{3\textrm{-sep}}   . \label{matrix3sep3}\enq  Note  that the
separability inequalities (\ref{a-bc3sep})-(\ref{matrix3sep3}) all
give bounds on anti-diagonal elements in terms of diagonal
elements.

 We will now show that these bounds  improve upon the  separability conditions  discussed  in section \ref{introsepconditions}.
We focus on the antidiagonal element $\rho_{1,8}$ (i.e., we
suppose that this is the largest antidiagonal matrix element)
since this is easiest for comparison. However, the same argument
holds for any other antidiagonal matrix element.

The D\"ur-Cirac conditions in terms of $|\rho_{1,8}|$ read as
follows. For partial separability under  the split $a$-$(bc)$: 
$2|\rho_{1,8}| \leq \rho_{4,4} +\rho_{5,5}$, under the split
$b$-$(ac)$: $2|\rho_{1,8}| \leq \rho_{3,3} +\rho_{6,6}$, and
lastly under the split $c$-$(ab)$: $2|\rho_{1,8}| \leq \rho_{2,2}
+\rho_{7,7}$. Next, the
Laskowski-\.Zukowski condition (\ref{antidiagonal})  gives for
$\rho \in  {\cal D}_3^{2\textrm{-sep}}$  that $|\rho_{1,8}|\leq
1/4$ and for $\rho \in {\cal D}_3^{3\textrm{-sep}} $ that
$|\rho_{1,8}|\leq1/8$. The fidelity condition (\ref{fidelity})
gives that if $\rho\in {\cal D}_3^{2\textrm{-sep}}$  then
$2|\rho_{1,8}|\leq \rho_{2,2} +\ldots+ \rho_{7,7}$.

In order to show that all these conditions are implied  by our
separability conditions, we employ some inequalities which hold
for all states $\rho$: $|\rho_{1,8}|^2\leq \rho_{1,1}\rho_{8,8}$
(this follows from (\ref{eenheid})), and $(\sqrt{\rho_{4,4}} -
\sqrt{\rho_{5,5}})^2\geq0\Longleftrightarrow2\sqrt{\rho_{4,4}\rho_{5,5}}\leq
\rho_{4,4} +\rho_{5,5}$, and similarly
$2\sqrt{\rho_{3,3}\rho_{6,6}}\leq \rho_{2,2} +\rho_{6,6}$ and
$2\sqrt{\rho_{2,2}\rho_{7,7}}\leq \rho_{2,2} +\rho_{7,7}$. Using
these trivial inequalities  one easily sees  that the conditions
(\ref{a-bc3sep})-(\ref{c-ab3sep}) imply the D\"ur-Cirac conditions
for separability under the three bipartite splits.  It is also
easy to see that the condition for 3-separability
(\ref{matrix3sep3}) strengthens the Laskowski-\.Zukowski condition
(\ref{antidiagonal}) for $k=3$. However, it is not so easy to see
 that (\ref{matrixbisep3}) strengthens both the fidelity and
Laskowski-\.Zukowski condition for $k=2$. We will nevertheless
show that this is indeed the case.

Let us use the symbols $\overset{A}{\leq}$ and
$\overset{2\textrm{-sep}}{\leq}$ to denote inequalities that  hold
for all states or for biseparable  states respectively. Combining
the above  trivial inequalities  with  condition
(\ref{matrixbisep3}) yields the following sequence of
inequalities: \beq
  4|\rho_{1,8}|- (\rho_{1,1}+\rho_{8,8}) \overset{A}{\leq}2 |\rho_{1,8}| \overset{2\textrm{-sep}}{\leq} 2\sqrt{\rho_{4,4}\rho_{5,5}} +2\sqrt{\rho_{3,3}\rho_{6,6}} +2\sqrt{\rho_{2,2}\rho_{7,7}}  \overset{A}{\leq}\rho_{2,2} + \cdots + \rho_{7,7}.
  \label{inequ3}
  \eeq
The inequality between the second and third expression is
(\ref{matrixbisep3}). It implies the other inequalities that
follow from (\ref{inequ3}). Comparing the first and fourth
expression of (\ref{inequ3}) one obtains the Laskowski-\.Zukowski
condition (\ref{antidiagonal}), while a comparison of the second
and fourth yields the fidelity criterion (\ref{fidelity}). Comparing the first and third term gives a 
new condition which was not previously mentioned. All these are implied by
condition (\ref{matrixbisep3}).

To end this section we show that the separability inequalities for $x=0$ give
Mermin-type separability inequalities \cite{mermin}. Consider the
Mermin operator for three qubits:
 \beq
  M^{(3)}:= X^{(1)}_a X^{(1)}_bY^{(1)}_c+ Y^{(1)}_aX^{(1)}_bX^{(1)}_c+
  X^{(1)}_aY^{(1)}_bX^{(1)}_c- Y^{(1)}_aY^{(1)}_bY^{(1)}_c,  \label{M3}
 \eeq
and define $M'^{(3)}$  in the same way, but with all
$X$ and $Y$ interchanged.   We can now use the
identity  $16( \av{{X^{(3)}_0}}^2 +\av{{Y^{(3)}_0}}^2)
=\av{M^{(3)}}^2+\av{M'^{(3)}}^2$ to obtain  from the  separability
conditions  (\ref{first3}) and (\ref{3sep3b})    the following quadratic
inequality for $k$-separability:
\begin{align}
   \label{quadratic3}
  16( \av{{X^{(3)}_0}}^2 +\av{{Y^{(3)}_0}}^2) =\av{M^{(3)}}^2+\av{M'^{(3)}}^2
   \leq 64\big(\frac{1}{4}\big)^k, ~~ \forall \rho \in {\cal
D}_3^{k\textrm{-sep}}.
   \end{align}
Of course, a similar bound holds when  $\av{X_0}^2 + \av{Y_0}^2$
in the left-hand side is replaced by   $\av{X_x}^2 + \av{Y_x}^2$
for $x=1,2,3$.
   This reproduces, for $N=3$,  the result  (\ref{quadraticN}) of Ref.~\cite{nagataPRL}.    From the density matrix representation, we
see that these Mermin-type separability conditions are in fact
equivalent to the Laskowski-\.Zukowski condition
(\ref{antidiagonal}).\forget{ when choosing the Pauli observables
for the set of local observables (for then one in effect obtains
the Laskowski-\.Zukowski condition for $k$-separability
\emph{simpliciter} $|\rho_{1,8}|^2\leq (1/4)^k$ which was shown to
be implied by our separability conditions). However, this holds
for any choice of orthogonal local observables since using a
suitable local unitary transformation one can transform the chosen
set of observables into the Pauli set.}\forget{These
 quadratic inequalities imply the following linear Mermin-type Bell-inequalities
 for partial separability: \beq |\av{M_3} | \leq 2^{3-k}, ~~ \forall \rho \in {\cal
D}_3^{k\textrm{-sep}}. \eeq These inequalities are sharp, i.e.\
$\sup_{\rho \in {\cal D}_3^{k\textrm{-sep}}} |\av{M_3} | =
2^{3-k}$. For full separability ($k=3$) this reproduces a result
obtained by Roy \cite{roy}, and for biseparability  ($k=2$) this
reproduces the result of \cite{tothseev}.} Note that these conditions do not distinguish the different classes
within level $k=2$, as was the case in \eqref{a-bc3sep}-\eqref{c-ab3sep}.

\subsection{$N$-qubit case}\label{Nqubitsection}

In this section we generalize the analysis of the previous section
to $N$ qubits to obtain conditions for $k$-separability  and
$\alpha_k$-separability. The proofs are analogous to the previous
cases, and will be omitted. Explicit conditions for
$k$-separability are presented for all levels $k = 1, \ldots, N$.
Further, we give a recursive procedure to derive
$\alpha_k$-separability conditions for each $k$-partite split
$\alpha_k$  at all level $k$. From these, one can easily construct
the conditions that distinguish all the classes in $N$-partite
separability classification by enumerating all possible logical
combinations of separability or inseparability under each of these
splits at a given level. We will however not attempt to write down
these latter  conditions explicitly since the number of classes
grows exponentially with the number of qubits.  We  start by
considering  bipartite splits, and biseparable states (level
$k=2$), and then move upwards to obtain separability conditions
for splits on higher levels.

We define $2^{(N-1)}$ sets of four observables
$\{X^{(N)}_x,~Y^{(N)}_x,~Z^{(N)}_x,~I^{(N)}_x \}$ , with  $x\in
\{0,1, \ldots,2^{(N-1)}-1\}$
 recursively from the
$(N-1)$-qubit observables:
\begin{align}
X^{(N)}_y &:=\frac{1}{2}\,(X^{(1)} \otimes X_{y/2}^{(N-1)} -Y^{(1)}\otimes Y_{y/2}^{(N-1)})
&X^{(N)}_{y+1} &:=\frac{1}{2}\,(X^{(1)} \otimes X_{y/2}^{(N-1)} +Y^{(1)}\otimes Y_{y/2}^{(N-1)})\nn
\\
Y^{(N)}_y  &:= \frac{1}{2}\,(Y^{(1)}\otimes X_{y/2}^{(N-1)}+X^{(1)} \otimes Y_{y/2}^{(N-1)} )
&Y^{(N)}_{y+1}  &:= \frac{1}{2}\,(Y^{(1)}\otimes X_{y/2}^{(N-1)}-X^{(1)} \otimes Y_{y/2}^{(N-1)} )\nn\\
Z^{(N)}_y  &:= \frac{1}{2}\,(Z^{(1)}\otimes I_{y/2}^{(N-1)}+I^{(1)} \otimes Z_{y/2}^{(N-1)} )
&Z^{(N)}_{y+1}  &:= \frac{1}{2}\,(Z^{(1)}\otimes I_{y/2}^{(N-1)}-I^{(1)} \otimes Z_{y/2}^{(N-1)} )\nn\\
  I^{(N)}_y  &:= \frac{1}{2}\,(I^{(1)} \otimes I_{y/2}^{(N-1)} +Z^{(1)}\otimes Z_{y/2}^{(N-1)})
    &I^{(N)}_{y+1}  &:= \frac{1}{2}\,(I^{(1)} \otimes I_{y/2}^{(N-1)}-Z^{(1)}\otimes Z_{y/2}^{(N-1)}),
  \label{Noperators}
  \end{align}
 with $y~\mathrm{even}$, i.e., $y\in \{0,2,4,\ldots\}$.   Analogous relations between these observables hold as those between the
  observables
  (\ref{set2})  and  (\ref{N3operators}). In particular, if the orientations of each triple  of local orthogonal
  observables is the same,  these sets form
  representations of the generalized
  Pauli group, and  every $N$-qubit state  obeys $\av{X^{(N)}_x}^2
+\av{Y^{(N)}_x}^2
  \leq\av{I^{(N)}_x}^2 -\av{Z^{(N)}_x}^2$,
  with equality only for pure states.

\subsubsection{Biseparability}

Consider a state that is separable under some bipartite split
$\alpha_2$ of the $N$ qubits. For each such split we get
$2^{(N-1)}$ separability inequalities in terms of the sets
$\{X^{(N)}_x,~Y^{(N)}_x,~Z^{(N)}_x,~I^{(N)}_x \}$ labeled by
$x\in\{0,1\ldots ,2^{(N-1)}-1\}$. These separability inequalities
provide necessary conditions for the $N$-qubit state to be
separable under the split under consideration. In order to find
these inequalities, we  first determine  the $N$-qubit  analogs of
the three-qubit pure state equalities (\ref{2_sep_a_bcX}) and
(\ref{2_sep_a_bcI}) corresponding to this bipartite split.  We
have not found a generic  expression  that lists them all for each
possible split and all $x$. However, for the split where the first
qubit is separated from the $(N-1)$ other qubits, i.e., $\alpha_2 =
a\textrm{-}(bc \ldots n)$ a generic form can be given:
\begin{align}
\av{X_x^{(N)} }^2 +\av{Y_x^{(N)} }^2&=
\frac{1}{4}\,(\,\av{X^{(1)}_a}^2+\av{Y^{(1)}_a}^2\,)\,
(\,\av{X^{(N-1)}_{x/2}}^2 +\av{Y^{(N-1)}_{x/2}}^2\,)=
 \av{X_{x+1}^{(N)} }^2 +\av{Y_{x+1}^{(N)} }^2  =\nn\\
 \av{I_x^{(N)} }^2 -\av{Z_x^{(N)} }^2&=
\frac{1}{4}\,(\,\av{I^{(1)}_a}^2-\av{Z^{(1)}_a}^2\,)\,
(\,\av{I^{(N-1)}_{x/2}}^2 -\av{Z^{(N-1)}_{x/2}}^2\,)=
 \av{I_{x+1}^{(N)} }^2 -\av{Z_{x+1}^{(N)} }^2,
 \label{partialeq}
\end{align}
where, without loss of generality, $x$ is chosen to be even, i.e.
$x\in\{0,2,4,\ldots\}$. For  other bipartite splits the sets of
observables labeled by $x$ are permuted, in a way depending on
the particular split.

For example,  for $N=4$  where $x\in \{0,1,\ldots,7\}$ the
equalities  (\ref{partialeq}) give the result for the split
$a$-$(bcd)$.  The corresponding equalities for  other bipartite splits are
obtained by the following permutations of $x$: for split $b$-$(acd)$:
$1 \leftrightarrow 3$ and $5 \leftrightarrow 7$; for split
$c$-$(abd)$: $1 \leftrightarrow 6$ and $3 \leftrightarrow
4$; and for split $d$-$(abc)$: $1 \leftrightarrow 4$ and $3
\leftrightarrow 6$. For the split $(ab)$-$(cd)$: $1
\leftrightarrow 2$ and $5 \leftrightarrow 6$; for  $(ac)$-$(bd)$:  $1 \leftrightarrow 7$ and $3
\leftrightarrow 5$; and lastly, for $(ad)$-$(bc)$:
$1 \leftrightarrow 5$ and $3 \leftrightarrow 7$.

For mixed states that are separable under a given bipartite split
the equalities (\ref{partialeq}) (and  their analogs obtained via
suitable permutations) become inequalities. We again state them
for
 the split $a$-$(bc\ldots n)$:
\begin{align} %\left.
\max \left\{ \begin{array}{c}
\av{X_x^{(N)} }^2 +\av{Y_x^{(N)} }^2
\\
\av{X_{x+1}^{(N)} }^2 +\av{Y_{x+1}^{(N)} }^2
\end{array} \right\}
\leq&
\min \left\{ \begin{array}{c}
\av{I_{x}^{(N)} }^2 -\av{Z_{x}^{(N)} }^2
\\
\av{I_{x+1}^{(N)} }^2 -\av{Z_{x+1}^{(N)} }^2
 \end{array}
 \right\}\leq \frac{1}{4}, ~~ \forall \rho \in {\cal
D}_N^{a\textrm{-}(bc\ldots n)} ~~~~\mbox{ with
$x\in\{0,2,4,\ldots\}$.}
 \label{partialeqM}
  \end{align}
The proof of (\ref{partialeqM}) is a straightforward
generalization of the convex analysis in section
\ref{twoqubitsection}. Again, for the other bipartite splits, the
 labels  $x$ are permuted in a way depending on the particular
split.

For a general biseparable state  $ \rho \in {\cal
D}_N^{2\textrm{-sep}}$, we thus obtain the following biseparability
conditions: \beq\label{Nk2first} \av{{X}_x^{(N)}}^2
+\av{{Y}_x^{(N)}}^2 \leq 1/4,~ \forall x, ~~~\forall \rho \in
{\cal D}_N^{2\textrm{-sep}} , \eeq which is equivalent to the
Laskowski-\.Zukowski condition for $k=2$ (as will be shown below).  And
just as in the three-qubit case, we also obtain a stronger condition
\beq\label{Nk2} \sqrt{\av{{X}_x^{(N)}}^2 +\av{{Y}_x^{(N)}}^2 }\leq
\sum_{y \neq x} \sqrt{\av{I_{y}^{(N)}}^2 -\av{Z_{y}^{(N)}}^2},
~~~\forall \rho \in {\cal D}_N^{2\textrm{-sep}} , ~~ \mbox{ with
$x,y=0,1,\ldots, 2^{(N-1)}-1$.}\eeq
 Violation of this inequality is a sufficient
condition for  full inseparability, i.e., for full $N$-partite
entanglement.

The inequalities (\ref{Nk2}) are stronger than the fidelity
criterion (\ref{fidelity}) and the Laskowski-\.Zukowski criterion
(\ref{antidiagonal}) for $k=2$,  and inequalities
(\ref{partialeqM}) are  stronger than the D\"ur-Cirac condition
(\ref{dccondition}) for separability under bipartite splits. This
will be shown below in subsection \ref{Nmatrix}. 

\subsubsection{Partial separability criteria for levels $2<k\leq N$}
For levels  $k>2$ we  sketch  a procedure to find
$\alpha_{k+1}$-separability inequalities recursively from
inequalities at the preceding level. Suppose that at level $k$ the
inequalities are given for separability under each $k$-partite
split $\alpha_k$ of the $N$ qubits, and  that these
$\alpha_k$-separability inequalities take the form:
 \beq
 \max_{ x \in z_i^{\alpha_k}}  \av{{X}_x^{(N)}}^2 +\av{{Y}_x^{(N)}}^2 \leq
 \min_{x \in z_i^{\alpha_k}} \av{I_x}^2 - \av {Z_x}^2
 \leq
\frac{1}{4^{(k-1)}},\forget{~~i=1,2,\ldots, 2^{(N-k)},} ~~ \forall
\rho \in {\cal D}_N^{\alpha_k}, ~~~~i\in\{1,2,\dots,2^{(N-k)}\}.
\label{recursivek} \eeq \forget{with
$\mathcal{Q}_{z_i^{\alpha_k}}$ the maximum of some set of
expressions $\av{{X}_y^{(N)}}^2 +\av{{Y}_y^{(N)}}^2$  and
$\mathcal{P}_{z_i^{\alpha_k}}$ the minimum of some set  of
expressions  $\av{I_{y}^{(N)}}^2 -\av{Z_{y}^{(N)}}^2$} where
$z_i^{\alpha_k}$ denote `solution sets' for the specific
$k$-partite split $\alpha_k$. \forget{This gives a total of
$2^{(N-k)}$ inequalities.   A solution set $z_i^{\alpha_k}$ is a
set of values for $x$ that are grouped together in a separability
inequality for the specific split $\alpha_k$.} For example, in the
case of three qubits,  the solution sets   for  the bipartite
split $a$-$(bc)$ are $z_1^{a\textrm{-}(bc)}=\{0,1\}$ and
$z_2^{a\textrm{-}(bc)}=\{2,3\}$, as can be seen from
(\ref{ineq_2sep}). The solution sets for other bipartite splits can be read off (\ref{ineq_2sepb}) and
(\ref{ineq_2sepc}) so as to give: $z_1^{b\textrm{-}(ac)}=\{0,3\}$, 
$z_2^{b\textrm{-}(ac)}=\{1,2\}$, and  $z_1^{c\textrm{-}(ab)}=\{0,2\}$, 
$z_2^{c\textrm{-}(ab)}=\{1,3\}$. And for future purposes we list them for the case of four qubits in table \ref{tabel1} below. These were obtained by  determining \eqref{partialeqM} for $N=4$ and for all bi-partite splits $\alpha_2$.
 \begin{table}[h]
\begin{centerline}{
\begin{tabular}{|c||c|c|c|c|c|c|c|}
\hline
split $\alpha_2$&$a$-$(bcd)$&$b$-$(acd)$&$c$-$(abd)$&$d$-$(abc)$&$(ab)$-$(cd)$&$(ac)$-$(bd)$&$(ad)$-$(bc)$\\
\hline\hline
$z_1^{\alpha_2}$&$\{ 0,1  \}$&$\{ 0 ,3  \}$&$\{ 0 ,6  \}$&$\{ 0 ,4  \}$&$\{0  ,2  \}$&$\{ 0 ,7  \}$&$\{  0, 5 \}$\\
$z_2^{\alpha_2}$&$\{  2,3  \}$&$\{ 1 ,2  \}$&$\{ 1 , 7 \}$&$\{1  ,5  \}$&$\{ 1 ,3  \}$&$\{  1,6  \}$&$\{ 1 ,4  \}$\\
$z_3^{\alpha_2}$&$\{  4,5  \}$&$\{  5,6  \}$&$\{  2,4  \}$&$\{ 2 ,6  \}$&$\{ 4 ,6  \}$&$\{ 2 ,5  \}$&$\{ 2 ,7  \}$\\
$z_4^{\alpha_2}$&$\{  6,7  \}$&$\{  4, 7 \}$&$\{  3, 5 \}$&$\{ 3 ,7  \}$&$\{ 5 ,7  \}$&$\{ 3 ,4  \}$&$\{ 3 ,6  \}$\\
\hline
\end{tabular}}
\end{centerline}
\caption{Solution sets for the seven different bi-partite splits of four qubits.} \label{tabel1}
\end{table}

\forget{Let us denote the total number of possible splits for an
$N$-partite state on level $k$ by $\sharp_k^N$
\cite{explicit_expression}.} Now move one level higher and
consider a given $(k+1)$-partite split $\alpha_{(k+1)}$. This
split  is contained in a total number of $\binom{k+1}{2}=k(k+1)/2$~
$k$-partite splits $\alpha_k$. Call the collection of these
$k$-partite splits $\mathcal{S}_{\alpha_{(k+1)}}$. We then obtain
preliminary  separability inequalities for the  split
$\alpha_{k+1}$ from the conjunction of all separability
inequalities for the splits $\alpha_k$ in the set
$\mathcal{S}_{\alpha_{(k+1)}}$.
 To be
specific, this yields:
 \beq\max_{ \alpha_k \in \mathcal{S}_{\alpha_{k+1}}}
\max_{ x \in z_i^{\alpha_k}}
 \av{{X}_x^{(N)}}^2 +\av{{Y}_x^{(N)}}^2\leq \forget{\min_{ x\in z_i^{\alpha_k}, \alpha_k\in\mathcal{S}_{\alpha_{(k+1)}}}}
\min_{ \alpha_k \in\mathcal{S}_{\alpha_{(k+1)}}} \min_{x \in
z_i^{\alpha_k}}   \av{I^{(N)}_x}^2 - \av{Z^{(N)}_x}
^2
   \leq
\frac{1}{4^{k-1}}, ~~ \forall \rho \in {\cal
D}_N^{\alpha_{(k+1)}}, \label{recursivek+1}
 \eeq
This may be written more compactly as \beq  \max_{ x \in
z_i^{\alpha_{k+1}}}
 \av{{X}_x^{(N)}}^2 +\av{{Y}_x^{(N)}}^2\leq \min_{x \in
z_i^{\alpha_{k+1}}}   \av{I^{(N)}_x}^2 - \av{Z^{(N)}_x} ^2
   \leq
\frac{1}{4^{k-1}}, ~~ \forall \rho \in {\cal D}_N^{\alpha_{(k+1)}}
\label{recursivek+1compact} ~~~~i\in\{1,2,\dots,2^{(N-k-1)}\}.
 \eeq
 (In fact, this can be regarded as an implicit definition of the
 solution sets  $z^{\alpha_{k+1}}_i$.)
 More importantly,   by
an argument similar to that leading from (\ref{3sep3}) to
(\ref{3sep3b}) one finds a stronger numerical bound in the utmost
right-hand side of these inequalities, namely $4^{-k}$ instead
of $4^{-(k-1)}$. Thus, the final result is: \beq  \max_{ x \in
z_i^{\alpha_{k+1}}}
 \av{{X}_x^{(N)}}^2 +\av{{Y}_x^{(N)}}^2\leq \min_{x \in
z_i^{\alpha_{k+1}}}   \av{I^{(N)}_x}^2 - \av{Z^{(N)}_x} ^2
   \leq
\frac{1}{4^{k}}, ~~ \forall \rho \in {\cal
D}_N^{\alpha_{(k+1)}},~~~~i\in\{1,2,\dots,2^{(N-k-1)}\}.
\label{recursivek+1final} \eeq This shows that the
$\alpha_k$-separability  inequalities  indeed take the same  form
as (\ref{recursivek}) at all levels.

As an example of this recursive procedure, take  $N=4$, set $k=3$,
and choose the split $a$-$b$-$(cd)$. This split is contained in
three $2$-partite splits $a$-$(bcd)$, $b$-$(acd)$ and
$(ab)$-$(cd)$.\forget{The separability inequalities for these splits have
the following solution sets (four per split):  $z_i^{a\textrm{-}(bcd)}$:
$\{0,1\},\{2,3\},\{4,5\},\{6,7\}$, $z_i^{b\textrm{-}(acd)}$:
$\{0,3\},\{1,2\}\{4,7\},\{5,6\}$, and $z_i^{(ab)\textrm{-}(cd)}$:
$\{0,2\},\{1,3\}\{4,6\},\{5,7\}$ respectively.} 
Using (\ref{recursivek+1})  and the first, second and fifth column of table \ref{tabel1}  
 one obtains the following two solutions sets  for the split  $a$-$b$-$(cd)$:    $z_1^{a\textrm{-}b\textrm{-}(cd)}=\{0,1,2,3\}$ and $z_2^{a\textrm{-}b\textrm{-}(cd)}=\{4,5,6,7\}$. This leads to the separability
inequalities:
\begin{align}\begin{array}{clcl}
 \max\limits_{x\in \{ 0,1,2,3\}} \av{X_{x}^{(4)} }^2 +\av{Y_{x}^{(4)}
}^2  &\leq&
 \min\limits_{x\in \{ 0,1,2,3\}} \av{I_{x}^{(4)} }^2 -\av{Z_{x}^{(4)}
}^2&
 \leq \frac{1}{16} \\
 \max\limits_{x\in \{ 4,5,6,7\}} \av{X_{x}^{(4)} }^2 +\av{Y_{x}^{(4)}
}^2 &\leq&
 \min\limits_{x\in \{ 4,5,6,7\}} \av{I_{x}^{(4)} }^2 -\av{Z_{x}^{(4)}
}^2&
 \leq \frac{1}{16}
 \end{array},  ~~ \forall \rho \in {\cal
D}_4^{a\textrm{-}b\textrm{-}(cd)}.
\label{43}
 \end{align}
For  other $3$-partite splits\forget{besides  the split
$a$-$b$-$(cd)$} the inequalities can be obtained in a similar way so as to give table \ref{tabel2} below.

\begin{table}[h]
\begin{centerline}{
\begin{tabular}{|c||c|c|c|c|c|c|} \hline
split $\alpha_3$ &  $a$-$b$-$(cd)$ & $(ab)$-$c$-$d$& $a$-$b$-$(cd)$&  $(ac)$-$b$-$d$& $(ad)$-$b$-$c$ & $(bd)$-$a$-$c$\\
\hline\hline $z_1^{\alpha_3}$ &  \{0,1,2,3\} &\{0,2,4,6\} &\{0,1,4,5\}
&\{0,3,4,7\} &\{0,3,5,6\} &\{0,1,6,7\} \\ $z_2^{\alpha_3}$ &
 \{4,5,6,7\} &\{1,3,5,7\} &\{2,3,6,7\}
&\{1,2,5,6\} &\{1,2,4,7\} &\{2,3,4,5\}\\\hline
\end{tabular}}
\end{centerline}
\caption{Solution sets for the six different $3$-partite splits of four qubits.} \label{tabel2}
\end{table}

As a special case, we mention the result for full separability,
i.e., for $k=N$. There is only one $N$-partite split, namely where
all qubits end up in a different set.  Further, there is only one
solution set $z_i^{\alpha_N}$ and it contains all $x \in \{0,1,
\ldots,2^{(N-1)}-1\}$. States $\rho$ that are separable under this
split thus obey: \beq \max_{x} \av{{X}_x^{(N)}}^2
+\av{{Y}_x^{(N)}}^2 \leq \min_{x} \av{I_{x}^{(N)}}^2
-\av{Z_{x}^{(N)}}^2 \leq \frac{1}{4^{(N-1)}}, ~~~\forall \rho \in
{\cal D}_N^{N\textrm{-sep}}. \label{NNsep} \eeq
 Violation of this inequality is a sufficient condition for
some entanglement to be present in the $N$-qubit state. The
condition (\ref{NNsep}) strengthens the Laskowski-\.Zukowski
condition  (\ref{antidiagonal}) for $k=N$ (to be shown below).

For an $N$-qubit $k$-separable state $ \rho \in {\cal
D}_N^{k\textrm{-sep}}$, i.e., a state that is a convex mixture of
states that are separable under some $k$-partite split,
 we  obtain from (\ref{recursivek+1final}) the following
$k$-separability  conditions: \beq\label{Nksepfirst}
\av{{X}_x^{(N)}}^2 +\av{{Y}_x^{(N)}}^2\leq
\frac{1}{4^{(k-1)}},~\forall x, ~~~\forall \rho \in {\cal
D}_N^{k\textrm{-sep}}, \eeq which is equivalent to the
Laskowski-\.Zukowski condition (\ref{antidiagonal}) for all $N$
and $k$ (this will be shown below using the density matrix
formulation of these conditions).  However, in analogy to
(\ref{2sep3}) we also obtain  the stronger condition: \beq
\sqrt{\av{{X}_x^{(N)}}^2 +\av{{Y}_x^{(N)}}^2}\leq \min_l  \sum_{y
\in \mathcal{T}_{k,l}^{N,x} }\sqrt {\av{{I}_y^{(N)}}^2
-\av{{Z}_y^{(N)}}^2 } ,~~~\forall \rho \in {\cal
D}_N^{k\textrm{-sep}}, \label{Nksep} \eeq   where, for given $N,
k$ and $x$, $\mathcal{T}_{k,l}^{N,x}$ denotes  a tuple  of values
of $y\neq x$, each one being picked  from each of the solutions
sets $z_i^{\alpha_{k}}$ that contain $x$, where $\alpha_k$ ranges
over all the $k$-partite splits of the $N$ qubits. In general,
there will be many ways of picking such values, and we use $l$ as
an index to label such tuples.

For example, in the case $N=3$, there are a total of 6 solution
sets (two for each of the three bipartite splits): $\{ 0,1\},
\{2,3\}, \{ 0,2\}, \{1,3\} ,  \{ 0,3\}, \{1,2\}$. If we set $x=0$
and pick a member different  from  0 from each of those sets that
contain $0$, we find: $\mathcal{T}_{2,1}^3= \{ 1,2,3\}$. This is
in fact the only such choice and thus $l=1$.  Thus, in this
example condition (\ref{Nksep}) reproduces the result
(\ref{2sep3}).

As a more complicated example, take $N=4$, $k=3$, and choose again
$x=0$. In this case there are six 3-partite splits each of which has two solution sets, as given in table \ref{tabel2}. 
 The solution sets that contain 0 are all on the top row of this
table. There are now many ways of constructing a tuple by picking
elements  that differ from 0 from  each of these sets , for
example ${\cal T}^{4,0}_{3,1}= \{ 1,2,1,3,3, 1\}$, ${\cal
T}^{4,0}_{3,2} = \{ 1,2,1,3,3,6\}$, etc. In this case one  has to
take a minimum in (\ref{Nksep}) over all these $l=1, \ldots, 3^6$
tuples.

\forget{
 The minimum is taken over all distinct sets
$l$. In each set $l$ the distinct occurrences of $y$ are picked
from a total number of sets that is equal to the total number
$\sharp_k^N$ \cite{explicit_expression} of possible splits
$\gamma_{k}$ that exist  for an $N$-partite state on this level
$k$.} \forget{ Violation of condition (\ref{Nksep}) is a
sufficient criterion for $(k)$-inseparability, i.e., for
$(k-1)$-separable entanglement (and thus for at least $
\integer{N/(k-1)}$-partite entanglement). {\bf niet
k-separabiliteit?}} For $k=2$, condition (\ref{Nksep}) reduces
to (\ref{Nk2}) and for $k=N$ to (\ref{NNsep}). For these values of
$k$, the condition is stronger than (\ref{Nksepfirst}) (see the
next section). For $k\neq 2,N$, this is still an open question.

To conclude this subsection, let us recapitulate.   We have found
separability conditions in terms of local orthogonal observables
for each of the $N$ parties that are  necessary for
$k$-separability  and for separability under  splits $\alpha_k$ at
each level on the hierarchic separability classification.
\forget{The criteria can distinguish all different levels from
each other by a bound that decrease by a factor of four for each
level. Furthermore, they allow for distinguishing the classes
within each level, using the finer characterization made possible
by the interconnections between splits at different levels as
captured by the notion of contained splits.} Violations of these
separability conditions give sufficient criteria for $k$-separable
entanglement and $m$-partite entanglement with $\integer{N/k}\leq
m\leq N-k+1$. The separability conditions are stronger than the
D\"ur-Cirac condition for separability under specific splits, and
stronger than the fidelity condition and the Laskowski-\.Zukowski
condition for biseparability. The latter condition is also
strengthened for $k=N$. These implications are shown in the next
section.\forget{The strength of the conditions will be further
commented on in section \ref{strengthsection}. We first take a
further look at the derived separability conditions.}

\subsubsection{The conditions in terms of matrix elements}\label{Nmatrix}
Choosing the Pauli matrices $\{\sigma_x^{(j)},
\sigma_y^{(j)},\sigma_z^{(j)}\}$ as local orthogonal observables,
with the same orientation at each qubit,    allows one to
formulate the separability conditions in terms of the density
matrix elements $\rho_{i,j}$ on the standard $z$-basis
\cite{basis}.
 For these choices we obtain:\begin{alignat}{2}
X_0^{(N)}&= \ket{0}\bra{1}^{\otimes N} +\ket{1}\bra{0}^{\otimes N},
&&\av{ X_0^{(N)}}=2\mathrm{Re}\, \rho_{1,d}, \nn\\
Y_0^{(N)}&= -i\ket{0}\bra{1}^{\otimes N} +i\ket{1}\bra{0}^{\otimes N}, \qquad
&&\av{Y_0^{(N)}}=-2\mathrm{Im}\, \rho_{1,d}, \nn\\
I_0^{(N)}&=  \ket{0}\bra{0}^{\otimes N} +\ket{1}\bra{1}^{\otimes N},
&&\av{ I_0^{(N)}}=\rho_{1,1}+ \rho_{d,d},\nn\\
Z_0^{(N)}&= \ket{0}\bra{0}^{\otimes N} -\ket{1}\bra{1}^{\otimes N},
&&\av{ Z_0^{(N)}}=\rho_{1,1} -\rho_{d,d}, \label{zbasisrel}
\end{alignat}
where $d= 2^N$.  Analogous relations hold for $X_x^{(N)},
~Y_x^{(N)},~ Z_x^{(N)},~I_x^{(N)}$ for $x\neq 0$.

Let us treat the case  $N=4$ in detail. First, consider the level
$k=2$. Biseparability under the split $a$-$(bcd)$ gives the
following inequalities for the anti-diagonal matrix elements:
\begin{align}
\label{n4k2}
\begin{array}{clcl}
 \max \{ |\rho_{1,16}|^2, |\rho_{8,9}|^2\} &\leq& \min\{ \rho_{1,1}\rho_{16,16}, \rho_{8,8}\rho_{9,9} \} &\leq1/16 \\
 \max \{|\rho_{2,15}|^2, |\rho_{7,10}|^2\} &\leq& \min\{\rho_{2,2}\rho_{15,15}, \rho_{7,7}\rho_{10,10} \} &\leq1/16\\
\max \{|\rho_{3,14}|^2, |\rho_{6,11}|^2\} &\leq& \min\{\rho_{3,3}\rho_{14,14}, \rho_{6,6}\rho_{11,11} \} &\leq1/16\\
 \max \{|\rho_{5,12}|^2, |\rho_{4,13}|^2\} &\leq&\min\{\rho_{5,5}\rho_{12,12}, \rho_{4,4}\rho_{13,13} \}&\leq1/16
 \end{array},~~~~ \mbox{ $ \forall \rho \in {\cal D}_4^{a\textrm{-}(bcd)}$}
\end{align}
 The analogous inequalities for separability under  other
bipartite splits are obtained  by suitable permutations on the
labels. Indeed,  for split $b$-$(acd)$ labels 8 and 5, 9 and 12, 2 and 3, 5 and 14 are permuted, which we denote as: $(8,9, 2,15)
\leftrightarrow ( 5, 12,3,14)$; and for split $c$-$(abd)$:
$(8,9,2,15)\leftrightarrow (3,14,5,12)$; for split
$d$-$(abc)$: $(8,9, 3,14) \leftrightarrow (2,15, 5,12)$; for
the split $(ab)$-$(cd)$: $(8,9,3,14 ) \leftrightarrow (4,13,
7,10)$; for  $(ac)$-$(bd)$: $(8,9, 5,12)  \leftrightarrow
(6,11, 7,10)$; and lastly, for the split $(ad)$-$(bc)$: $(8,9,
5,12) \leftrightarrow (7,10, 6,11)$. For a general biseparable
state we obtain \beq
 |\rho_{1,16}| \leq \sqrt{\rho_{2,2}\rho_{15,15}} +
\sqrt{\rho_{3,3}\rho_{14,14}}+ \ldots+\sqrt{\rho_{8,8}\rho_{9,9}},
 ~~~ \forall \rho \in {\cal D}_4^{2\textrm{-sep}}
,\eeq
and analogous for the other anti-diagonal elements.

Next, consider one level higher, i.e., $k=3$.  There are six
different $3$-partite splits  for a system consisting of four
qubits. For separability under each such split a different set of
inequalities can be obtained from (\ref{recursivek+1}). To be more
precise,  such a set consists of the conjunction of all the
separability inequalities for the bipartite splits at level $k=2$
this particular $3$-partite split is contained in. For $N=4$ each
$3$-partite split is contained in three bipartite splits. For
example, for separability under split $a$-$b$-$(cd)$
 we obtain:
\begin{align}\begin{array}{cccl}
\max
\{ \begin{array}{l}  |\rho_{1,16}|^2,
|\rho_{8,9}|^2,
|\rho_{4,13}|^2,
|\rho_{5,12}|^2
        \end{array}\}
      &  \leq&\min
        \{ \begin{array}{l} \rho_{1,1}\rho_{16,16},~\rho_{8,8}\rho_{9,9},~
                    \rho_{4,4}\rho_{13,13},~
                    \rho_{5,5}\rho_{12,12}
        \end{array}
        \}&\leq1/64.
       \\
      \max
\{ \begin{array}{l}  |\rho_{2,15}|^2,
|\rho_{3,14}|^2,
|\rho_{6,11}|^2,
|\rho_{7,10}|^2
        \end{array}\}
       & \leq&\min
\{ \begin{array}{l} \rho_{2,2}\rho_{15,15},~\rho_{3,3}\rho_{14,14},~
                    \rho_{6,6}\rho_{11,11},~
                    \rho_{7,7}\rho_{10,10}
        \end{array}\} &\leq1/64
        \end{array}, ~ \forall \rho
\in {\cal D}_4^{a\textrm{-}b\textrm{-}(cd)}.
        \end{align}
      This is the density matrix formulation of (\ref{43}).

A general $3$-separable state  $\rho \in {\cal
D}_4^{3\textrm{-sep}}$ is a convex mixture of states that each are
separable under some such $3$-partite split. The separability
condition follows from (\ref{Nksep}): \beq\label{MelementsN4k3}
 |\rho_{1,16}|\leq \min_{l}
(\sum_{j\in\tilde{\mathcal{T}}_{3,l}^{4,0}}
\sqrt{\rho_{j,j}\rho_{17-j,17-j} }), ~~~\forall \rho\in {\cal
D}_4^{3\textrm{-sep}}, \eeq
 where $\tilde{\mathcal{T}}_{3,l}^{4,0}$  is the tuple of indices $j\in\{1,16\}$ that label the anti-diagonal density matrix elements $\rho_{j,17-j}$ corresponding to the density matrix formulation of the set of operators  $\av{{X}_y^{(4)}}^2 +\av{{Y}_y^{(4)}}$ with $y$ determined by 
 $\mathcal{T}_{3,l}^{4,0}$. Here we have used that the anti-diagonal element $\rho_{1,16}$ corresponds to   $\av{{X}_0^{(4)}}^2 +\av{{Y}_0^{(4)}}^2$. 
              For  $N=4$, $k=3$ there are six possible splits, so for each $l$, $j$ is picked from a total of six sets.
                For the case under consideration the sets are $\{1,4,5,8 \}, \{1,2,3,4 \}, \{1,3,5,7 \}, \{1,2,5,6 \}, \{1,2,7,8 \}$, and $\{1,3,6,8 \}$. For each $l$ one chooses  a tuple  of values of $j$ where one value is picked from each of these six sets, except for the value $1$ which is excluded.    Analogous inequalities are obtained for the other anti-diagonal matrix elements.

Finally for full separability ($k=4$) we get: \beq
\max\{|\rho_{1,16}|^2, |\rho_{2,15}|^2,\ldots,
|\rho_{8,9}|^2\}\leq \min \{  \rho_{1,1}\rho_{16,16},
\rho_{2,2}\rho_{15,15}, 
\ldots,\rho_{8,8}\rho_{9,9}   \}\leq 1/256, ~~~ \forall \rho \in
{\cal D}_4^{4\textrm{-sep}}.       \eeq
For general $N$, it is easy to see that (\ref{Nk2first}) yields
the Laskowski-\.Zukowski condition (\ref{antidiagonal}).
 It is instructive to look
at the extremes of biseparability  and full separability, since
for them explicit forms can be  given. For $k=2$ condition
(\ref{Nk2}) reads: \beq |\rho_{{l},\bar{l}}|\leq \sum_{n \neq
l,\bar{l}}\sqrt{\rho_{n,n}\rho_{\bar{n},\bar{n}}}/2 ,~~ \forall \rho \in
{\cal D}_N^{2\textrm{-sep}}   ~~~\mbox{where $\bar{l}=d+1-l$,
$\bar{n}=d+1-n$}, ~~~l,n\in \{1,\dots,d\}.  \label{N2gen} \eeq
\forget{
 For example, for $l=1$ this gives
 \begin{align}
 \label{Nmatrixelements}
|\rho_{1,d}|\leq \sqrt{\rho_{2,2}\rho_{d-1,d-1}}+
\sqrt{\rho_{3,3}\rho_{d-2,d-2}}+\ldots+\sqrt{\rho_{d/2,d/2}\rho_{d/2+1,d/2+1}},~~
\forall \rho \in {\cal D}_N^{2\textrm{-sep}}. 
\end{align}} For  $k=N$, we
can reformulate condition (\ref{NNsep}) as \beq
 \max\{|\rho_{1,d}|^2, |\rho_{2,d-1}|^2\ldots         \}
 \leq\min\{  \rho_{1,1}\rho_{d,d}, \rho_{2,2}\rho_{d-1,d-1}, \ldots
        \}\leq1/4^{N},~~ \forall \rho \in {\cal D}_N^{N\textrm{-sep}} .\label{matrixfull}
\eeq  It is easily seen that the condition (\ref{matrixfull}) is
stronger than the Laskowski-\.Zukowski condition
(\ref{antidiagonal}) for this case.

Again, these inequalities  give bounds on anti-diagonal matrix
elements in  terms of diagonal ones on the $z$-basis.
 These density matrix representations depend on the choice of the Pauli matrices as the local observables.
 However, every other triple of  locally orthogonal observables with the same orientation can
 be obtained from the Pauli matrices by suitable local basis transformations, and therefore this matrix
 representation does not loose generality.
 Choosing different orientations of the triples  one obtains the corresponding inequalities by
  suitable permutations of anti-diagonal matrix elements.

We will now show that (\ref{N2gen}) is indeed stronger than the
fidelity condition (\ref{fidelity}) and the Laskowski-\.Zukowski
condition (\ref{antidiagonal}) for $k=2$ by following the same
analysis as in the three-qubit case. We again assume, for
convenience, that the antidiagonal element $\rho_{1,d}$ is the
largest of all antidiagonal elements. Using some inequalities that
hold for all states together with the condition (\ref{N2gen}) for
biseparability  we get the following sequence of inequalities for
 $\rho_{1,d}$: \beq
  4|\rho_{1,d}|- (\rho_{1,1}+\rho_{d,d}) \overset{A}{\leq}2 |\rho_{1,d}| \overset{2\textrm{sep}}{\leq} 2\sqrt{\rho_{2,2}\rho_{d-1,d-1}} + \cdots+ 2\sqrt{\rho_{d/2,d/2}\rho_{d/2+1,d/2+1}}  \overset{A}{\leq}\rho_{22} + \cdots + \rho_{d-1,d-1}.
  \label{stronginequ}
  \eeq
The inequality in the middle is (\ref{N2gen}). It implies all
other inequalities in the sequence (\ref{stronginequ}). The
inequality between the first and fourth term yields the
Laskowski-\.Zukowski condition for $k=2$, and between the second
and fourth gives the fidelity criterion in the formulation
(\ref{equivfidelity}). One also sees that the fidelity criterion
is stronger than the Laskowski-\.Zukowski condition for $k=2$.

We finally discuss two examples showing that the biseparability
 condition (\ref{N2gen}) is stronger in detecting full
entanglement than other methods.
 First, consider the family of $N$-qubit
states \beq\label{states} \rho_N'=
\lambda_0^+\ket{\psi_0^+}\bra{\psi_0^+}
+\lambda_0^-\ket{\psi_0^-}\bra{\psi_0^-}+
\sum_{j=1}^{2^{N-1}-1}\lambda_j(\ket{\psi_k^+}+\ket{\psi_j^-})(\bra{\psi_j^+}+\bra{\psi_j^-}).
\eeq The states (\ref{states})  violate (\ref{N2gen}) for all
$|\lambda_0^+ - \lambda_0^-|\neq 0$ and are thus detected as fully
entangled by that condition.   In that case they are also
inseparable under any split. The fidelity criterion
(\ref{equivfidelity}), however, detects these states as
fully entangled only for
 $|\lambda_0^+ - \lambda_0^-|\geq \sum_j \lambda_j$.
Violation of (\ref{N2gen}) thus allows for detecting more  states
of the form $\rho_N'$ as fully entangled than violation of the
fidelity criterion. Further, the D\"ur-Cirac criteria detects these
states  as inseparable under any split  for $|\lambda_0^+ -
\lambda_0^-|> 2\lambda_j$, $\forall j$, which includes less states
than a violation of (\ref{N2gen}).
 This generalizes the observation of Ref.\
\cite{ota} from two qubits to the $N$-qubit case.

Secondly, consider the $N$-qubit GHZ-like states
$\ket{\theta}=\cos{\theta}\ket{0}^{\otimes
N}+\sin{\theta}\ket{1}^{\otimes N}$    \forget{with density matrix
 \beq\label{thetastate}
\ket{\theta}\bra{\theta}=
 \left (
\begin{array}{lll}
\cos^2{\theta} &\cdots&\cos{\theta}\sin{\theta}\\
~~\vdots&&~~\vdots\\
\cos{\theta}\sin{\theta} &\cdots&\sin^2{\theta}\\
\end{array}\right ).
 \eeq}
 We can easily read off from the density matrix $\ket{\theta}\bra{\theta}$
 that the far off-antidiagonal matrix elements $\rho_{1,d}=\rho_{d,1}$ is equal
 to $\cos{\theta}\sin{\theta}$  and that the diagonal  matrix elements
 $\rho_{2,2}, \ldots, \rho_{d-1,d-1}$ are all equal to zero.
  Using (\ref{N2gen}) we see that these states are
fully $N$-partite entangled for $\rho_{1,d}=\cos{\theta}\sin{\theta}\neq 0$,
  i.e., for all $\theta\neq 0,\pi/2$ (mod $\pi$). Thus, all fully
entangled states of this form are detected by condition
(\ref{N2gen}), including those not  detectable by any standard
multipartite Bell inequality \cite{zukow2002}.

\subsubsection{Relationship to Mermin-type inequalities for partial separability and LHV models}\label{LHVmermin}
We will now show that the separability inequalities of the
previous section imply already known Mermin-type inequalities
\cite{mermin} for partial separability.

Using the identity $ 2^{(N+1)}(\av{X^{(N)}_0}^2
+\av{Y^{(N)}_0}^2)=\av{M^{(N)}}^2 +\av{M'^{(N)}}^2$, for the Mermin
operators (\ref{merminN})  together with the upper bound for the
separability inequality  of (\ref{Nksepfirst}) for $x=0$ gives the
following sharp quadratic inequality: 
\beq \label{quadraticN2}
\av{M^{(N)}}^2 +\av{M'^{(N)}}^2\leq2^{(N+3)}\big(\frac{1}{4}\big)^k,~~
\forall \rho \in {\cal D}_N^{k\textrm{-sep}}. 
\eeq 
which
reproduces the  result (\ref{quadraticN}) found by
\cite{nagataPRL}. Since (\ref{Nk2first}) is equivalent to
(\ref{antidiagonal}) we see that the Mermin type separability
condition is in fact one of Laskowski-\.Zukowski conditions
written in terms of local observables $X$ and $Y$. \forget{ These
quadratic inequalities
  straightforwardly give rise to sufficient conditions for
  $k$-separable  entanglement, i.e., if
 $\av{M^{(N)}}^2 +\av{M'^{(N)}}^2 >2^{(N+3)}({1}/{4})^{(k+1)}$,
  then the $N$-partite state $\rho$ is $k$-separable entangled and not
 \mbox{$(k+1)$}-separable entangled,
 and at least $m$-partite entangled, with $m\geq \integer{N/k}$.
If the state is separable according to some single split where the largest split contains exactly $m'$
qubits then it is a sufficient condition for
$m'$-partite entanglement.

 Furthermore, the quadratic inequalities  (\ref{quadraticN}) imply the following sharp linear Mermin-type
inequality  for $k$-separability: \beq %\label{linearN}
|\av{M^{(N)}}|\leq 2^{(\frac{N+3}{2})}\big(\frac{1}{2}\big)^k,~~
\forall \rho \in {\cal D}_N^{k\textrm{-sep}}. \eeq \forget{This
inequality  is sharp: $\sup_{\rho \in {\cal D}_N^{k\textrm{-sep}}}
|\av{M_N} | = 2^{(N+3-2k)/2}$. }If (\ref{linearN}) is violated the
$N$ qubit state $\rho$ is $(k-1)$-separably
 entangled and it has at least $m$-partite entanglement,
 with $m\geq \integer{N/(k-1)}$. For $k=N$ inequality (\ref{linearN})  reproduces a
 result obtained by Roy \cite{roy}.}

As a special case we consider a split of the form
$\{1\},\ldots,\{\kappa\},\{\kappa+1,\ldots,n\}$. Any state that
is separable under this  split is $(\kappa +1)$-separable so we
get the condition $\av{M^{(N)}}^2 +\av{M'^{(N)}}^2\leq
2^{(N-2\kappa+1)}$, and hence  $|\av{M^{(N)}}| \leq
2^{(N-2\kappa+1)/2}$. This strengthens the result of Gisin and
Bechmann-Pasquinucci \cite{gisin} by a factor $2^{\kappa/2}$ for
these specific Mermin operators (\ref{merminN}).

\begin{figure}[!h]
\includegraphics[scale=0.9]{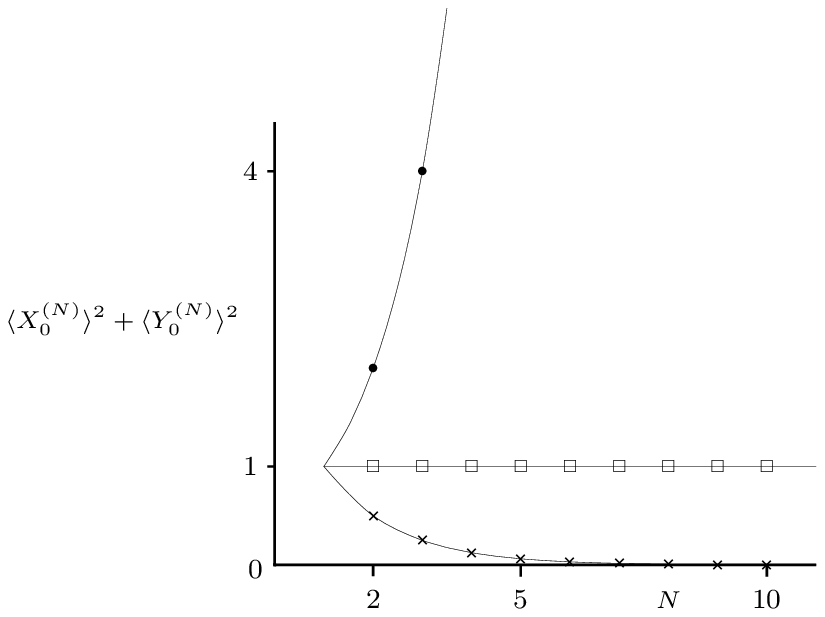}
\caption{The maximum value for $\av{X_0}^2 + \av{Y_0}^2$
obtainable by entangled quantum states (dots), by separable
quantum states (crosses) and by LHV models (squares), plotted as a
function of the number of qubits $N$.  Note the exponential
divergence between both the maxima obtained for entangled states
as well as for separable states compared to the LHV value, where
the former maximum is exponentially increasing and the latter
maximum is exponentially decreasing.} \label{grafiek1}
\end{figure}

As another special case of the inequalities (\ref{quadraticN2}),
consider $k=N$. In this case, the inequalities express a condition
for full separability of $\rho$. These inequalities are maximally
violated by fully entangled states by an exponentially increasing
factor of $2^{N-1}$, since the maximal value of $|\av{M^{(N)}}|$ for
any quantum state $\rho$ is $2^{(N+1)/2}$ \cite{wernerwolf}.
  Furthermore,  LHV models violate them also by an exponentially increasing factor of $2^{(N-1)/2}$, since
for all $N$, LHV models allow a maximal value for $|\av{M^{(N)}}|$  of
$2$ \cite{gisin,seevuff}, which is  a factor $2^{(N-1)/2}$ smaller
than the quantum maximum using entangled states. This bound for LHV models is sharp since
the maximum is attained by  choosing  the LHV expectation
values $\av{\sigma_x^i}=\av{\sigma_y^i}=1$ for all $i \in \{1,
\ldots, N\}$.   This shows that there are exponentially increasing
gaps between the values of  $|\av{M^{(N)}}|$ attainable by fully separable states,
fully entangled states and LHV models. This is shown in Figure \ref{grafiek1}.

That the maximum violation of multipartite Bell inequalities allowed
by quantum mechanics grows exponentially with $N$ with respect to
the value obtainable by LHV models has been known for quite some
years \cite{mermin,wernerwolf}. However, it is
equally remarkable that the maximum value obtainable by separable
quantum states \emph{exponentially decreases} in comparison  to the
maximum value obtainable by LHV models, cf.\  Fig.\ \ref{grafiek1}.
 We thus see exponential divergence between separable quantum
states and LHV theories:  as $N$ grows,  the latter are able to give
correlations that need more and more entanglement in order to be
reproducible in quantum mechanics.

But why does quantum mechanics have correlations larger than those
obtainable by a LHV model? Here we give an argument showing that
it is not the degree of entanglement but the degree of
inseparability that is responsible. The degree of entanglement of
a state may be quantified by the value $m$ that indicates the
$m$-partite entanglement of the state, and the degree of
inseparability by the value of $k$ that indicates the
$k$-separability of the state. Now suppose we have $100$ qubits.
For partial separability of $k\geq 51$ no state of these $100$
qubits can violate the Mermin inequality (\ref{linearN}) above the
LHV bound, although the state could be up to $50$-partite
entangled ($m\leq50$). However, for $k=2$, a state is possible
that is also $50$-partite entangled, but which  violates the
Mermin inequality by an exponentially large factor of $2^{97/2}$.
For $k<N$, a $k$-separable state is always entangled in some way,
so we see that it is the degree of partial separability, not the
amount of entanglement  in a multi-qubit state that determines the
possibility of a violation of the Mermin inequality. Of course,
some entanglement must be present, but the inseparability aspect
of the state determines the possibility of a violation. This is
also reflected in the fact that for a given $N$ it is the value of
$k$, and not that of $m$, which determines the sharp upper bounds
of the Mermin inequalities.

\section{Experimental strength of the conditions for $k$-separable entanglement detection}\label{strengthsection}
  Violations of the above
conditions for partial separability  provide sufficient
criteria for detecting $k$-separable entanglement (and $m$-partite
entanglement with $\integer{N/k}\leq m\leq N-k+1$). It has already
been shown that these criteria are stronger than the Laskowski-\.Zukowski
criterion for  $k$-inseparability  for $k=2,N$ (i.e., detecting
some and full entanglement), the fidelity criterion for full
inseparability (i.e., full entanglement) and the D\"ur-Cirac
criterion for inseparability under splits. In this section we will
elaborate further on the experimental usefulness and strength of
these entanglement criteria, when focusing on specific $N$-qubit states. The strength of an
entanglement criterion to detect a given entangled state may be
assessed by determining how well it copes with two desiderata
\cite{tothguhne2}:  the noise robustness of the criterion for this given state should be high, and the number of local measurements
settings needed for its implementation should be small.

In this section we will first take a closer look at the issue of
noise robustness and at the number of required settings for
implementation of the separability criteria, both in the general
state-independent case and in the case of detecting target states.
We then show the strength of the criteria for a variety of
specific $N$-qubit states.\forget{ In particular, we show (i)
detection of bound entanglement for $N\geq 3$, (ii) detection of
$N$-qubit $W$-states with great noise robustness, (iii) detection
of a six qubit cluster state and four qubit Dicke state, (iv)
great noise and decoherence robustness in detecting the $N$-qubit
GHZ state and better noise robustness than the stabilizer witness
criterion for detecting these latter states.}

\subsection{Noise robustness and the number of  measurement settings}

White noise robustness  of an entanglement criterion for a given
entangled state is the maximal fraction $p_0$ of  white noise which may be admixed to this state  so that the state can no longer be detected as entangled by the
criterion. Thus, for a given entangled state $\rho$, the
noise robustness of a criterion is the threshold value $p_0$ for which the state
$\rho=p\,\1/2^N +(1-p) \rho$, with $p\geq p_0$ can
no longer be detected by that criterion.

So, for the criterion for detecting full entanglement
(\ref{N2gen}), the white noise robustness is found by solving the
threshold equation for $p_0$: \beq |(1-p_0) \rho_{l,\bar{l}}| =
\sum_{j\neq l} \sqrt{(\frac{p_0}{2^N}
+(1-p_0)\rho_{j,j})(\frac{p_0}{2^N}
+(1-p_0)\rho_{\bar{\jmath},\bar{\jmath}})}, \label{noiseeq} \eeq
\forget{This equation is quadratic in $p_0$ and can be easily solved. For
$l$ that gives the maximum of $|\rho_{l,\bar{l}}|- \sum_j
\sqrt{\rho_{j,j}\rho_{\bar{\jmath},\bar{\jmath}}},  j\neq l$.} The
state is fully entangled for $p<p_0$.

For the criterion (\ref{matrixfull}), for detecting some entanglement,
  one finds a similar threshold equation: \beq \max_l \{|(1-p_0)
\rho_{l,\bar{l}}|^2\} = \min_j \{(\frac{p_0}{2^N}
+(1-p_0)\rho_{j,j})(\frac{p_0}{2^N}
+(1-p_0)\rho_{\bar{\jmath},\bar{\jmath}})\}.
 \label{noiseeqsep}
\eeq This equation is quadratic and easily solved. Again, the state
is entangled for $p<p_0$.

A local measurement setting
\cite{setting,terhalComputSci,guhnehyllus}  is an observable such
as $\mathcal{M}=\sigma_1\otimes\sigma_l \ldots\otimes \sigma_N$,
where $\sigma_l$ denote single qubit  observables for each of the
$N$ qubits. Measuring such a setting (determining all coincidence
probabilities of the $2^N$ outcomes) also enables one to determine
the probabilities for observables like $\1\otimes\sigma_2
\ldots\otimes \sigma_N$, etc.\ \cite{guhne2007}.  Now consider the
observables $X_x^{(N)}$ and $Y_x^{(N)}$ that appear in the
separability criteria of (\ref{partialeq})-(\ref{Nksep}). As it is 
easily seen from their definitions in (\ref{Noperators}), one can
measure such an observable   using $2^N$  local settings. However,
  these  same $2^N$ settings then suffice to measure the observables $X_x^{(N)}$ and
$Y_x^{(N)}$ for all other $x$ since these are linear combinations
of the same settings. Thus,  $2^N$ measurement settings are
sufficient to determine $\av{{X}_x^{(N)}}$ and $\av{{Y}_x^{(N)}}$
for all $x$. It remains to determine the number of settings needed
for the terms $\av{I_{x}^{(N)}}$ and $\av{Z_{x}^{(N)}}$. For all
$x$ these terms contain only two single-qubit observables:
$Z^{(1)}$ and $I^{(1)}= \1$. They can thus be measured by a single
setting, i.e.,  $\left (Z^{(1)}\right)^{\otimes N}$.

 Thus, in total $2^N+1$ settings are needed in order to test the separability conditions. This number
grows exponentially  with the number of qubits. However, this is
the price we pay for being so general, i.e., for having criteria
that work for all states.  If we apply the criteria to detecting
forms of inseparability and entanglement of specific entangled
$N$-qubit states, this number can be greatly reduced. Knowledge of
the target state enables one to select a single separability
inequality for an optimal value of $x$ in
(\ref{partialeq})-(\ref{Nksep}). Violation of this single
inequality is then sufficient for detecting the entanglement in
this state, and, as we will now show, the required number of
settings then grows only linear in $N$, with $N+1$ being the
optimum for many states of interest.

For simplicity, assume  that the local observables featuring in
the criteria are the Pauli spin observables with the same
orientation for each qubit.
  We can then readily use the density matrix representations of the separability
criteria given at the end of each subsection in the previous
section. Choosing the local observables  differently amounts to
performing suitable bases changes to the density matrix
representations and would not affect the argument.

The  matrix representations  of the conditions show that only some
anti-diagonal matrix elements and the values of some diagonal
matrix elements have to be determined in order to test whether
these inequalities are violated. Indeed, observe that for all $x$~
$\av{I_{x}^{(N)}}^2-\av{Z_{x}^{(N)}}^2=4\rho_{j,j}\rho_{
\bar{\jmath}, \bar{\jmath}}$ with $\bar{\jmath}=d+1-j$ for some $j
\in \{1,2,\ldots,d\}$ and $\av{X_{x}^{(N)}}^2 -\av{Y_{x}^{(N)}}
^2=4|\rho_{j ,\bar{\jmath}}|^2$ denotes some
anti-diagonal matrix element.
 It suffices to consider $x=0$ since conditions for other values of $x$
are obtained by some local unitary basis changes that
will be explicitly given later on. We now want to rewrite the
density matrix representation for this single separability
inequality with $x=0$ in terms of less than $2^N+1$ settings.

Determining the diagonal matrix elements requires only a single
setting, namely $\sigma_z^{\otimes N}$.  Next,  we should
determine the modulus of the far-off anti-diagonal element $\rho_{1,d}$
($d=2^N$) by measuring
  $X_0^{(N)}$
   and
$Y_0^{(N)}$, since $\av{X_0^{(N)}} =2$Re$\rho_{1,d}$ and
$\av{Y_0^{(N)}} =2$Im$\rho_{1,d}$ (cf. (\ref{zbasisrel})).
Following the method of  \cite{guhne2007}, these matrix elements
can be obtained from two settings $\mathscr{M}_l$  and
$\tilde{\mathscr{M}}_l$, given by
\begin{align}\label{real}
\mathscr{M}_l&=\big(   \cos(\frac{l \pi}{N})\sigma_x + \sin(\frac{l
\pi}{N})\sigma_y \big)^{\otimes N},~~~
l=1,2,\ldots,N~,\\
\label{imaginary0} \tilde{\mathscr{M}}_l&= \big( \cos(\frac{l
\pi+\pi/2}{N})\sigma_x +\sin( \frac{l \pi+\pi/2}{N})\sigma_y
          \big)^{\otimes N},~~~ l=1,2,\ldots,N.
\end{align}
These  operators obey:
\begin{align}
\sum_{l=1}^N (-1)^l\, \mathscr{M}_l&=   N\, X_0^{(N)}, \label{real2}\\
 \sum_{l=1}^N (-1)^l\,\tilde{\mathscr{M}}_l&= N\, Y_0^{(N)}. \label{imaginary}
\end{align}
The proof of (\ref{real2}) is given in \cite{guhne2007} and
(\ref{imaginary}) can be proven in the same way.

These relations show that the imaginary and the real part of an
anti-diagonal element can  be determined by the $N$ settings
$\mathscr{M}_l $ and $\tilde{\mathscr{M}}_l$ respectively.  This
implies that the biseparability  condition (\ref{N2gen}) needs
only $2N+1$ measurement settings. However, if each anti-diagonal
term is  real valued (which is often the case for states of
interest) it can be determined by the $N$ settings $\mathscr{M}_l
$, so  that in total $N+1$ settings suffice.

Implementation of the criteria for other $x$ involves determining
the modulus of  some other anti-diagonal matrix element instead of
the far-off anti-diagonal element $\rho_{1,d}$. The settings that
allow for this  determination can be obtained from  a local
unitary rotation on the settings $\mathscr{M}_l$ and
$\tilde{\mathscr{M}}_l$ needed to measure $|\rho_{1,d}|$. This can
be done as follows.

Suppose we want to determine the modulus of the matrix element
$\rho_{j,\bar{\jmath}}$. The unitary rotation to be applied is
given by $
U_j=\sigma_{j_1}\otimes\sigma_{j_2}\otimes\ldots\otimes\sigma_{j_N}$
with $j=j_1 j_2\ldots j_N$ in binary notation, with $\sigma_0=\1$
and $\sigma_1=\sigma_x$.  The settings that suffice are then given
by $\mathscr{M}_{j,l}=U_j \,\mathscr{M}_l\,U_j^{\dagger}$  and
$\tilde{\mathscr{M}}_{j,l}
=U_j\,\tilde{\mathscr{M}}_l\,U_j^{\dagger}$  ($l=1,2,\ldots, N$).
For example,  take  $N=4$ and suppose we want to determine
$\rho_{5,4}$. We obtain the required settings by applying the
local unitary $U_5=\1\otimes\sigma_x\otimes \1\otimes\sigma_x$
(since the binary notation of $5$ on four bits is $0101$) to the
two settings $\mathscr{M}_l $ and $\tilde{\mathscr{M}_l}$  given
in (\ref{real}) and (\ref{imaginary}) respectively that for $N=4$
allow for determining $|\rho_{1,16}|$. In conclusion, using the
above procedure the modulus of each anti-diagonal element can be
determined using $2N$ settings, and in case they are  real (or
imaginary) $N$ settings suffice.

Since the strongest separability inequality for the specific
target state  under consideration is chosen, this reduction in the
number of settings does not reduce the noise robustness for
detecting forms of entanglement  as compared to that obtained
using the entanglement criteria  in terms of the usual settings
$X_x^{(N)}$, etc.  \forget{In fact, the white noise robustness
$p_0$ for detecting full entanglement is determined via
(\ref{N2gen}) by solving the following biseparability
\emph{simpliciter} threshold equation for $p_0$: \beq |(1-p_0)
\rho_{l,\bar{l}}| = \sum_{j\neq l} \sqrt{(\frac{p_0}{2^N}
+(1-p_0)\rho_{j,j})(\frac{p_0}{2^N}
+(1-p_0)\rho_{\bar{\jmath},\bar{\jmath}})}, \label{noiseeq} \eeq
\forget{This equation is quadratic in $p_0$ and can be easily
solved.} for $l$ that gives the maximum of $|\rho_{l,\bar{l}}|-
\sum_j \sqrt{\rho_{j,j}\rho_{\bar{\jmath},\bar{\jmath}}},  j\neq
l$. The state is fully entangled for $p<p_0$.

For detecting some entanglement  the white noise robustness $p_0$ is
determined via (\ref{matrixfull}) by solving the following $N$-separability \emph{simpliciter} threshold
equation for $p_0$:
\beq
\max_l \{|(1-p_0) \rho_{l,\bar{l}}|^2\} =
\min_j \{(\frac{p_0}{2^N} +(1-p_0)\rho_{j,j})(\frac{p_0}{2^N} +(1-p_0)\rho_{\bar{\jmath},\bar{\jmath}})\}.
 \label{noiseeqsep}
\eeq This equation is quadratic in $p_0$ and can be easily solved.
The state is entangled for $p<p_0$. }

 In conclusion, if the state
to be detected is known, the $2N$ settings of  (\ref{real}) and
(\ref{imaginary0})  together with the single setting
$\sigma_z^{\otimes N}$ suffice, and in case this state has solely
real or imaginary anti-diagonal matrix elements only $N +1$
settings are needed. The white noise robustness using these
settings  is just as great as using the general condition that use
the observables $X_x^{(N)}$ and $Y_x^{(N)}$, and is found by
solving (\ref{noiseeq}) or (\ref{noiseeqsep}) for detecting full
and some entanglement respectively.

As a final note, we observe that in order to determine the modulus
of not just one but of all anti-diagonal matrix elements it is
more efficient to use the observables $X_x^{(N)}$, $Y_x^{(N)}$
than the observables of (\ref{real}) and (\ref{imaginary0}). The
first method needs $2^N$ settings to do this and the second needs
$2^NN/2$ settings (since  there are $2^N/2$ independent
anti-diagonal elements), i.e.,  the latter needs more settings
than the former for  all $N$.

\forget{\subsubsection{examples}
 We will next apply the above procedure to two examples.
 The celebrated $N$-qubit GHZ state $\ket{\Psi_{\textrm{GHZ},0}^N}=
 (\ket{0^{\otimes N}}+\ket{1^{\otimes N}})/\sqrt{2}$
has only two non-zero off diagonal matrix elements $\rho_{1,d}$ and $\rho_{d,1}$
which are real valued and thus equal to each other. This state can be detected as
fully entangled using the criterion
$(\rho_{1,d})^2>\max\{ \rho_{j,j}\rho_{\bar{\jmath},\bar{\jmath}} \}$,  $j\in \{2,\dots,d-1\}$.  As
discussed above, for its implementation it  needs only
$N+1$  measurement settings ($N$ settings $\mathscr{M}_l$ given in (\ref{real}) plus the setting $
\sigma_z^{\otimes N}$).  The criterion has noise robustness of $p_0=1/(1+2^{(1-N)})$,
which goes to $1$ for large $N$. (In section \ref{GHZexample} this state is further analyzed
and other criteria for detecting it as entangled are also considered).
}

Let us apply the above procedure to an example, taken from  Ref.\
\cite{guhne2007},  the so-called four-qubit singlet state, which
is\forget{a state in a decoherence free subspace (i.e., invariant
under a simultaneous unitary rotation on all qubits). It is} given
by: \beq\label{4singlet} \ket{\Phi_4}=(\ket{0011} +\ket{1100}
-\frac{1}{2}(\ket{01} +\ket{10}) \otimes(\ket{01}+\ket{10}       )
)/\sqrt{3}. \eeq For detecting it as fully entangled
(\ref{noiseeq}) gives a noise robustness $p_0=12/29\approx0.41$,
and for detecting it as entangled (\ref{noiseeqsep}) gives a noise
robustness of $16/19\approx 0.84$. The implementation needs
$16+1=17$ settings.

 This number of settings can be reduced by using the fact that this state has
only real anti-diagonal matrix elements and that we need only look
at the largest anti-diagonal element. As shown above, this matrix
element can be measured in $4$ settings. Thus the total number of
settings required is reduced to only $5$.  The off-diagonal matrix
element to be determined is $\ket{0011} \bra{1100}$. The four
settings that allow for this determination are obtained from the
four settings given in (\ref{real}) by applying the unitary
operator $U_3=\1\otimes\1\otimes\sigma_x\otimes \sigma_x$ to these
settings.

For comparison, note that in Ref.\ \cite{guhne2007} it was shown
that the so-called projector-based witness for the state
($\ref{4singlet}$) detects full entanglement with a white noise robustness $p_0=0.267$ and
uses $15$ settings, whereas  the optimal witness from
\cite{guhne2007} uses only  $3$ settings and has $p_0=0.317$. Here
we obtain $p_0\approx0.41$ using $5$ settings, implying a
significant  increase in  white noise robustness using only two
settings more.

This  example gives the largest noise robustness when the
conditions are measured in the standard $z$-basis.  However, sometimes one obtains larger noise
robustness when the state is first rotated so as to be expressed
in a different basis before it is analyzed.  For example, consider
the four qubit Dicke state $\ket{2,4}$, where
$\ket{l,N}=\binom{N}{l}^{-1/2}\sum_k
\pi_k(\ket{1_1,\ldots,1_l,0_{l+1},\ldots,0_N})$ are the symmetric
Dicke states \cite{dicke}  (with $\{ \pi_k (\cdot)\}$  the set of
all distinct permutations of the $N$ qubits).  In the standard
basis this state does not violate any of the separability
conditions we have discussed above. However, if each qubit is
rotated around the $x$-axis by $90$ degrees all of the
separability conditions can be violated with quite high noise
robustness. Indeed, it is detected as inseparable under all splits
through violation of conditions (\ref{partialeqM}) for
$p<p_0=16/19\approx0.84$ and as fully entangled through violation
of condition (\ref{Nk2}) for $p<p_0=4/11\approx0.36$ using $5$
settings. For comparison, Chen {\it et al.} \cite{chen}  used
specially constructed entanglement witnesses for detection of full
entanglement in these states, and they obtained  as noise
robustness $p_0=2/9\approx 0.22$ using only $2$ settings.  We have not
performed an optimization procedure, so it is unclear whether or
not the values obtained for $p_0$ can be improved.

\subsection{Noise and decoherence robustness for the $N$-qubit GHZ state}\label{GHZexample}
In this subsection  we  determine the robustness of our
separability criteria for detecting the $N$-qubit GHZ state in
five kinds of noise processes (admixing white and colored noise,
and three types of decoherence: depolarization, dephasing and
dissipation of single qubits). 
We give the noise robustness as a function
of $N$ for detecting some entanglement, inseparability with respect to all splits and  full entanglement.
 We  compare the results for white noise robustness of the
criteria for full entanglement to that of the fidelity criterion
(\ref{fidelitycriterion}) and to that of the so called stabilizer
criteria of Refs. \cite{tothguhne2,tothguhne3}.

 The $N$-qubit GHZ state
$\ket{\Psi_{\mathrm{GHZ},0}^N}
=\frac{1}{\sqrt{2}}(\ket{0}^{\otimes N} +\ket{1}^{\otimes N}$) can
be transformed into a mixed state $\rho_N$ by  admixing noise to
this state or by decoherence. Let us consider the following five
such processes.

(i) Mixing in a fraction $p$ of white noise  (also called 'generalized Werner states' \cite{pittenger}) gives:
\beq\label{whitenoiseghz}
\rho_N^{\textrm{(i)}}=(1-p)\ket{\Psi_{\mathrm{GHZ},0}^N}\bra{\Psi_{\mathrm{GHZ},0}^N} +p\frac{\1}{2^N}.
\eeq

(ii) Mixing in a fraction $p$ of colored noise \cite{noise2} gives:
\beq
\rho_N^{\textrm{(ii)}}= (1-p)\ket{\Psi_{\mathrm{GHZ},0}^N}\bra{\Psi_{\mathrm{GHZ},0}^N} +\frac{p}{2}
(\ket{0\ldots0}\bra{0\ldots0} +\ket{1\ldots1}\bra{1\ldots1}).
\eeq

(iii) A depolarization process \cite{noise} with a
depolarization degree $p$ of a single qubit gives: 
\begin{align}
\forget{\ket{i}\bra{j}&\longrightarrow (1-p)\ket{i}\bra{j}
+p\,\delta_{ij} \frac{\1}{2}, \nn \\}
\rho_N^{\textrm{(iii)}}=\frac{1}{2}\big{[}  \big{(}
(1-\frac{p}{2}) \ket{0}\bra{0} + \frac{p}{2}\ket{1}\bra{1}
\big{)}^{\otimes N}
+\big{(}&\frac{p}{2}\ket{0}\bra{0} +  (1-\frac{p}{2})\ket{1}\bra{1}\big{)}^{\otimes N}\nn\\
&+(1-p)^N\big{(}\ket{0}\bra{1}^{\otimes N}+\ket{1}\bra{0}^{\otimes N}\big{)}\big{]}.
\end{align}

(iv) A dephasing process \cite{noise} with a dephasing
degree $p$ of a single qubit gives:  \begin{align}
\forget{\ket{i}\bra{j}&\longrightarrow &(1-p)\ket{i}\bra{j}
+p\,\delta_{ij} \ket{i}\bra{j}, \nn\\} \rho_N^{\textrm{(iv)}}  =
 \frac{1}{2} \big{[} \ket{0}\bra{0}^{\otimes
N}+\ket{1}\bra{1}^{\otimes N}+ (1-p)^N(\ket{0}\bra{1}^{\otimes
N}+\ket{1}\bra{0}^{\otimes N}) \big{]}. \end{align}

 (v) A dissipation process \cite{noise} with a dissipation degree $p$ of a single qubit (where
the ground state is taken to be $\ket{0}$) gives: \begin{align}
\forget{
\ket{i}\bra{i} & \longrightarrow & (1-p)\ket{i}\bra{i} +  p\ket{0}\bra{0},  \nn\\
\ket{i}\bra{j} & \longrightarrow & (1-p)^{1/2}\ket{i}\bra{j},
~~~~i\neq j ,    \nn \\} \rho_N^{\textrm{(v)}}  =  \frac{1}{2}
\big{[} \ket{0}\bra{0}^{\otimes N}+(p\ket{0}\bra{0}+
(1-p)\ket{1}\bra{1})^{\otimes N}+
(1-p)^{N/2}(\ket{0}\bra{1}^{\otimes N}+\ket{1}\bra{0}^{\otimes N})
\big{]}. 
\end{align}
We now consider the question for what values of $p$ these states $\rho_N^{(\textrm{i})}$ to $\rho_N^{(\textrm{v})}$ are detected as (i) containing some entanglement by the condition \eqref{NNsep},  and (ii) inseparable under any split by the conditions of the form \eqref{partialeqM} for all bipartite splits. 
 In other words, we determine the noise (or
decoherence) robustness of violations of all these conditions for $\rho_N^{(\textrm{i})}$ to $\rho_N^{(\textrm{v})}$.
We find the following threshold values $p_0$.
 \begin{alignat}{2}
 \label{GHZsomeN}
&\textrm{(i)}&& p_0= \frac{1}{1+2^{(1-N)}}, \nn
\\
&\textrm{(ii)}&& p_0 =1,~~~~ \forall N,\nn
\\
&\textrm{(iii)}~~&&  (1-p_0)^N= 
(1-\frac{p_0}{2})^{\alpha}(\frac{p_0}{2})^{(N-\alpha)} +(1-\frac{p_0}{2})^{(N-\alpha)}(\frac{p_0}{2})^{\alpha},\\
&\textrm{(iv)}&&p_0 =1,~~~~\forall N,\nn
\\
&\textrm{(v)}&& p_0= 1,~~~~\forall N,\nn 
\end{alignat} 
For cases (i), (ii), (iv) and (v) the threshold values $p_0$ for detecting some entanglement and inseparability with respect to all splits are the same because for these cases the product of the diagonal matrix elements  $\rho_{j,j}\rho_{\bar{\jmath},\bar{\jmath}}$ is the same for all $j\neq 1,d$.
 Only in case (iii) is this product different for different $j$.
We then have to take the minumum and maximum value, respectively, from which it follows that 
 $\alpha$ is to be set to $[N/2]$ for detecting some entanglement and to $1$ for detecting inseparability with respect to all splits. Here $[N/2]$ is the largest integer smaller or equal to $N/2$.

The result in case (i)
is in accordance with the results of Ref.\ \cite{duer2,duer},
where it is furthermore shown that the opposite holds as well,
i.e., iff $p<1/(1+2^{(1-N)})$ then $\rho_N^{\textrm{(i)}}$ is
inseparable under any split  and otherwise it is fully separable. Thus all states of the form
(\ref{whitenoiseghz}) that are inseparable under any split are
detected by violations of  the conditions of the form (\ref{partialeqM}) for all bipartite splits. The same holds for cases (ii),
(iv) and (v), since all states $\rho_N^{\textrm{(ii)}}$,
$\rho_N^{\textrm{(iv)}}$ and $\rho_N^{\textrm{(v)}}$ are
inseparable under any split for all $p<1$. In other words, as soon
as a fraction of the GHZ state is present, these states
 are inseparable under any split. In case
 (i) $p_0$  increases  monotonically from $p_0=2/3$ for $N=2$ to
$p_0=1$ for large $N$. For process (iii) these  limiting values
are not so straightforward: $p_0=(3-\sqrt{3})/3\approx0.42$ for
$N=2$, and $p_0=(5-\sqrt{5})/5\approx0.55$ for  large $N$. In
conclusion, the noise and decoherence robustness is high for all
$N$, except maybe for case (iii).

Next, consider the noise robustness for detecting full entanglement
by means of  the biseparability condition (\ref{Nk2}). The result is the
following: 
\begin{alignat}{2}
\label{GHZfullN} 
&\textrm{(i)}&& p_0=1/(2(1-2^{-N})),
\nn
\\
&\textrm{(ii)}&& p_0 =1,~~~~ \forall N,\nn
\\
&\textrm{(iii)}~~&& p_0 \approx 0.42,0.28,0.22,0.18,~~ N=2,3,4,5.
\\
&\textrm{(iv)}&&p_0=1,~~~~\forall N,\nn
\\
&\textrm{(v)}&& p_0\approx 1,0.48,0.39,0.35, ~~N=2,3,4,5.\nn 
\end{alignat}

For case (i) the noise robustness is equivalent to the fidelity
criterion (\ref{fidelitycriterion}). For large $N$  $p_0$ decreases  to the limit value $p_0=1/2$.
Case (ii) and (iv) have $p_0=1$, thus as soon as the states
$\rho_N^{\textrm{(ii)}}$ and $\rho_N^{\textrm{(iv)}}$ are
entangled they are fully entangled. For cases  (iii) and (v) we listed the
noise robustness found numerically for $N=2$ to $N=5$. These values decrease for increasing $N$.

Let us  compare the results for white noise robustness (case (i))
to the results  obtained from the so-called stabilizer formalism.
This formalism \cite{gottesman}  is used by T\'oth \& G\"uhne to
derive entanglement witnesses \cite{tothguhne2,tothguhne3} that are especially useful for minimizing the number of  settings
required to detect either full or some entanglement. Here we will
only consider the criteria formulated for detecting entanglement
of the $N$-qubit GHZ states. 
 The stabilizer witness by T\'oth \& G\"uhne that detects some
entanglement has $p_0=2/3$, independent of  $N$, and requires only
three settings (cf.\  Eq. (13) in \cite{tothguhne2}). The
strongest witness
 for full entanglement  of  T\'oth \& G\"uhne has a robustness
 $p_{0}=1/(3-2^{(2-N)})$ and requires only two settings  (cf.\ Eq. (23) in \cite{tothguhne2}).

\begin{figure}[!h]
\includegraphics[scale=1.0]{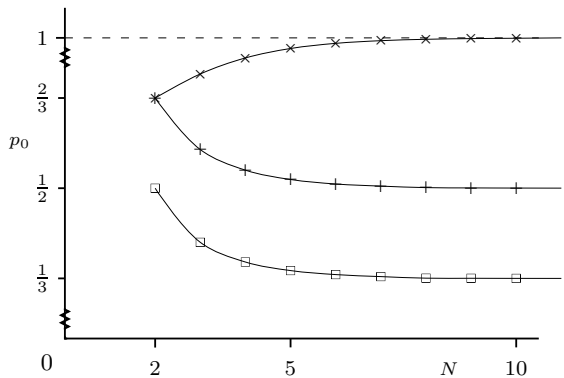}
\caption{ The threshold noise ratios $p_0$ for detection of full
$N$-qubit entanglement when admixing white noise to the $N$-qubit
GHZ state for the criterion (\ref{Nk2}) derived here (plus-signs)
and  for the stabilizer witness of Ref.\ \cite{tothguhne2}
 (squares). The noise robustness for
detecting  inseparability  under all splits as given in (i) in
(\ref{GHZsomeN}) is
 also plotted (crosses).
}
\label{grafiek4}
\end{figure}

Figure \ref{grafiek4} shows these threshold noise ratios for
detecting full entanglement for these three criteria. Note that
the criterion of T\'oth \& G\"uhne \cite{tothguhne2} needs only
two measurement settings, whereas our criteria need $N+1$
settings. So although the former are less robust against white
noise admixture, they compare favorably with respect to minimizing
the number of measurement settings.

Although we give a criterion for full entanglement that is
generally stronger than the fidelity criterion, for the
$N$-partite GHZ state this does not lead to  better noise
robustness. It appears that for large $N$ the noise threshold
$p_0=1/2$ is the best one can do. However, in the limit of large
$N$ the GHZ state is inseparable under all splits for all $p_0<1$,
as was shown in (i) in (\ref{GHZsomeN}). See also Figure
\ref{grafiek4}.
Furthermore, we have seen that if the state $\rho_N^{\textrm{(i)}}$   (i.e., the GHZ state with a fraction $p$ of white noise) is entangled it is also inseparable under any split. 
 Because of the high symmetry of both the GHZ state and white noise, one might
 conjecture that if the state $\rho_N^{\textrm{(i)}}$ is entangled it is also  fully entangled.
 At present, however it is unknown whether this is indeed true.
Detecting the states $\rho_N^{\textrm{(i)}}$ as fully entangled
appears to be a much more demanding task than detecting them as
inseparable under all splits. In the first case, for large $N$,
only a fraction of $50 \%$ noise is permitted,  in the second case
one can permit any noise fraction  (less than $100 \%$).
Note that we have given explicit examples of states that are diagonal in GHZ basis (cf. \eqref{dccondition} of section \ref{introsepconditions}), and that are inseparable under any split, but not fully entangled. But these are not of the form  $\rho_N^{\textrm{(i)}}$.

Lastly, we mention that  our criteria detect the various forms of entanglement and inseparability also if the state $\ket{\Psi_{\mathrm{GHZ},0}^N}$ is replaced by any other maximally entangled state (i.e., any state of the GHZ basis, cf. \eqref{DuerStates}), a feature which is not possible using linear entanglement witnesses.  There is no single linear witness  that detects entanglement of all maximally entangled states.

\subsection{Detecting bound entanglement for $N\geq 3$}
 Violation of the separability inequality (\ref{NNsep}) allows for detecting all bound
 entangled states of Ref.\ \cite{dur}. These states have the form
\beq \label{boundstate}
\rho_B=\frac{1}{N+1} \left( \ket{\Psi_{{\rm
GHZ},\alpha}^N}\bra{\Psi_{{\rm GHZ},\alpha}^N}
+\frac{1}{2}\sum_{l=1}^{N} P_l +\bar{P}_l \right),
 \eeq with $P_l$ the projector on the state
$\ket{0}_1\ldots\ket{1}_l \ldots \ket{0}_N$, and where $\bar{P}_l$
is obtained from $P_l$ by replacing all zeros by ones and vice
versa. For $N\geq4$ these states are entangled and have positive
partial transposition (PPT) with respect to transposition of any
qubit. This means they are  bound entangled \cite{bound}. Note
that they are detected as entangled  by  the $N$-partite Mermin
inequality $|M_N|\leq2$ of section \ref{Nqubitsection} only for
$N\geq8$ \cite{dur}. However, the condition (\ref{NNsep}) detects
them as entangled  for $N\geq 4$.  Thus all bound entangled states
of this form are detected as entangled by this latter condition.
\forget{, cf.\cite{nagata2007}. }The white noise robustness for
this purpose is $p_0=2^N/(2+2 N+ 2^N)$, which for $N=4$ gives
$p_0=8/13\approx0.615$ and  goes to $1$ for large $N$.
 Note that  for $N=4$, this state violates  the condition for
$4$-separability, and the condition for $3$-separability
(\ref{Nksep}), but not  the condition for $2$-separability. It is
thus at least $2$-separable entangled. It is not detected as fully
entangled by these criteria. (Of course, it could still be fully
entangled since these criteria are only sufficient and not
necessary for entanglement).  For general $N$ we have not
investigated the $k$-separable entanglement of the states
(\ref{boundstate}), although this can be readily performed using
the criteria of (\ref{Nksep}).

Another interesting bound entangled state is the so-called four
qubit Smolin state \cite{smolin} \beq \rho_S=
\frac{1}{4}\sum_{j=1}^4
\ket{\Psi^j_{ab}}\bra{\Psi^j_{ab}}\otimes\ket{\Psi^j_{cd}}\bra{\Psi^j_{cd}},
\eeq where $\{ \ket{\Psi^j} \}$ is the set of four Bell states $\{
\ket{\phi^{\pm}}, \ket{\psi^{\pm}} \}$, and $a,b,c,d$ label the
four qubits. This state is also detected as entangled by the
criterion (\ref{NNsep}), and with white noise robustness  $p_0=2/3$.
The Smolin state violates the separability conditions
(\ref{partialeqM}) for biseparability under the splits
$a$-$(bcd)$, $b$-$(acd)$, $c$-$(abd)$, $d$-$(abc)$. However, it is
separable under the splits $(ab)$-$(cd)$, $(ac)$-$(bd)$,
$(ad)$-$(bc)$ (cf.\ \cite{smolin}). This state is thus inseparable under
splits that partition the system into two subsets with one and
three qubits, but it is separable when each subset contains two
qubits.

So far we have detected bound entanglement for $N\geq 4$.  What
about $N=3$? Consider the three-qubit bound entangled state of
\cite{bound3}: \beq \rho=
\frac{1}{3}\ket{\Psi_{GHZ,0}^3}\bra{\Psi_{GHZ,0}^3}
+\frac{1}{6}(\ket{001}\bra{001} +\ket{010}\bra{010}
+\ket{101}\bra{101} +\ket{110}\bra{110} ). \eeq This state is
detected as entangled by the criterion (\ref{3sep3}), with white
noise robustness  $p_0= 4/7\approx0.57$. It violates the
biseparability condition (\ref{ineq_2sep}) for the split
$a$-$(bc)$ so it is at least biseparable entangled, but does not
violate the condition (\ref{2sep3}) for biseparability
 i.e., it is not detected as fully entangled.
In fact, it can be shown using the results of Ref. \cite{duer}
that this state is separable under the splits $b$-$(ac)$ and
$c$-$(ab)$.

\section{Discussion}\label{discussion}

We have discussed partial separability of quantum states by
distinguishing $k$-separability $\alpha_k$-separability and used
these distinctions to extend the classification proposed by D\"ur
and Cirac. We discussed the relationship of partial separability
to multipartite entanglement and distinguished the notions of a
$k$-separable entangled state and a $m$-partite entangled state
and indicated the interrelations of these kinds of entanglement.

 Next, we have presented necessary conditions for partial separability
  in the  hierarchic separability classification. These are  formulated
     in terms of experimentally accessible correlation inequalities for operators
     defined by products of local orthogonal observables.
   Violations of these inequalities provide, for all $N$-qubit states, criteria for
   the entire hierarchy of $k$-separable
entanglement, ranging from the levels $k$=1 (full or genuine
$N$-particle entanglement) to $k=N$ (full separability, no
entanglement), as well as for specific classes within each level.
Choosing the  Pauli matrices as the locally orthogonal observables
provided matrix representations of the criteria that bound
anti-diagonal matrix elements in terms of diagonal ones.

Further, the $N$-qubit Mermin-type separability inequalities for
partial separability were shown to follow from the partial
separability conditions derived in this paper. The biseparability
 conditions are stronger than the
fidelity criterion and the Laskowski-\.Zukowski criterion, and the
latter criterion is also shown to be strengthened for full
separability and biseparability. For separability under splits the conditions are
stronger than the D\"ur-Cirac conditions. Violation of these
conditions thus give entanglement criteria that  detect more
entangled states than violations of these three other separability conditions. \forget{
We therefore believe these state-independent entanglement criteria
to be the strongest experimentally accessible conditions for
multi-qubit entanglement applicable to all multi-qubit states.}

We have furthermore shown that the required number of measurement
settings for implementation of these criteria, which is $2^N +1$
in  general, can be drastically reduced if entanglement of a given
target state is to be detected. In  that case, it may be reduced
to  $2N+1$, and for multiqubit states with either real or
imaginary anti-diagonal matrix elements, only $N+1$ settings are
needed.

When comparing the entanglement criteria to other state-specific
multiqubit entanglement criteria it was found that the white noise
robustness was high for a great variety of interesting multiqubit
states, whereas  the number of required settings was only $N+1$.
However, these other state-specific entanglement criteria need
less settings although for the states analyzed here they give
lower noise robustness. Analyzing some specific target states
shows that the entanglement criteria detect bound entanglement for
$N\geq3$.

Furthermore, we applied  the entanglement criteria  for some and
full entanglement to  the $N$-qubit GHZ state subjected to two
different kinds of noise and three different kinds of decoherence.
The robustness against colored noise and against dephasing turns out
to be maximal (i.e., $p_0=1$) both for detecting some and full
entanglement.  It is remarkable that for large $N$ the GHZ state
allows for maximal white noise robustness for the state to remain
inseparable under all possible splits, whereas for detecting full
entanglement  the best known result -- to our best knowledge --
only allows for a white noise robustness of $p_0=1/2$. It would be very
interesting to search for full entanglement criteria that  can
close this gap, or if this is shown to be impossible to understand
why this is the case.

Orthogonality of the local observables is  crucial in the above
derivation of separability conditions. It is due to this
assumption that the multiqubit operators  form representations of
the generalized Pauli group. It would be interesting to analyze
the role of orthogonality in deriving the inequalities. For two
qubits it has been shown \cite{tradeoff} that when  orthogonality
is relaxed the separability conditions become less strong, and we
conjecture the same holds for their multiqubit analogs. Relaxing
the requirement of orthogonality has the advantage that some
uncertainty in the angles may be accommodated, which is desirable
since in real experiments it may be hard to measure perfectly
orthogonal observables.

It is also interesting that the separability inequalities are
equivalent to bounds on anti-diagonal matrix elements in terms of
products of diagonal ones.  We thus gain a new perspective on why
they allow for entanglement detection: they probe the values of
anti-diagonal matrix elements, which encode entanglement
information about the state; and if these elements are large
enough, this entanglement is detected.
 Note furthermore that compared to the Mermin-type separability inequalities
we need not do much more to obtain our stronger inequalities. We
must solely determine some diagonal matrix elements, and this can
be easily performed using the single extra setting
$\sigma_z^{\otimes N}$. It is also noteworthy that the comparison to the Mermin-type
separability inequalities shows that the strength of the
correlations allowed for by separable states is exponentially
decreasing when compared to the strength of the correlations
allowed for by LHV models.

Our recursive definition of the multipartite correlation operators
 (see (\ref{Noperators})) is by no means unique. One can generate
many new inequalities by choosing the locally orthogonal
observables differently, e.g.,  by permuting their order  in each
triple of local observables. It could well be that combining such
new inequalities with those presented here yield even stronger
separability conditions, as is indeed the case for pure two-qubit
states, cf.\ \cite{uffseev}. Unfortunately, we have no conclusive
answers  for this open question.

We end by suggesting three further lines of future research.
Firstly, it would be interesting to apply the entanglement
criteria to an even larger variety of $N$-qubit states than
analyzed here, including for example all $N$-qubit graph and Dicke
states. Secondly, the generalization from qubits to qudits (i.e.,
$d$-dimensional quantum systems)
 would, if indeed possible, prove very useful since strong partial separability criteria for $N$ qudits
have -- to our knowledge -- not yet been obtained. And finally, it would be beneficial to have optimization
procedures for choosing the set of local orthogonal observables featuring in the
entanglement criteria that gives the highest noise robustness for a given set of states.
We believe we have chosen such optimal sets for the variety of states analyzed here, but since no rigorous optimization was performed, our choices could perhaps be improved.

\section*{Acknowledgements}
We thank G\'eza T\'oth and Otfried G\"uhne for fruitful discussions on the strength of multipartite
entanglement criteria.


\begin{thebibliography}{99}
\bibitem{voet}For example, consider the following construction (inspired by Ref. \cite{tothguhne1}) where we use 
the $N$-qubit states $\ket{\psi_1}=\ket{e}_{12}\ket{0}_3\ket{0}_4
\cdots \ket{0}_N$; $\ket{\psi_2}=\ket{0}_1\ket{e}_{23}\ket{0}_4
\cdots\ket{0}_N$; \ldots,
$\ket{\psi_N}=\ket{0}_2\ket{0}_3 \cdots \ket{e}_{N,1}$, where $\ket{e}_{ij}$ is any entangled pure state of the two
parties $i$ and $j$ (mod $N$). Then the state $\rho =\sum_{i=1}
^N\ket{\psi_i}\bra{\psi_i}/N$  is inseparable under all splits, yet  by
construction $(N-1)$-separable.
\bibitem{basis}In the standard $z$-basis, $\rho_{i,j}=\bra{i'}\rho\ket{j'}$ with $i'=i-1$, $j'=j-1$
and where $i'=i_1i_2\ldots i_N$ and $j'=j_1j_2\ldots j_N$ are  in
binary notation. For example, for $N=4$: $\rho_{1,16}
=\bra{0000}\rho\ket{1111}$   and
$\rho_{8,11}=\bra{1000}\rho\ket{1011}$.
\end{thebibliography}
\end{document}